\documentclass[aps,pre,twocolumn,superscriptaddress,nobibnotes,nodoi,noeprint]{revtex4-2}
\usepackage{amsmath,amssymb,physics,bm,siunitx}
\usepackage{xcolor,graphicx}
\usepackage{soul}
\usepackage[colorlinks=true,citecolor=blue,urlcolor=blue,linkcolor=black]{hyperref}
\usepackage{setspace}

\definecolor{darkred}{rgb}{0.7, 0, 0} 

\begin{document}
	
\title{Phase locking and fractional Shapiro steps in collective
  dynamics of microparticles}

\author{Seemant Mishra} \email{semishra@uos.de}
\affiliation{Universit\"{a}t Osnabr\"{u}ck, Fachbereich
  Mathematik/Informatik/Physik, Barbarastra{\ss}e 7, D-49076
  Osnabr\"uck, Germany}

\author{Artem Ryabov} \email{rjabov.a@gmail.com} \affiliation{Charles
  University, Faculty of Mathematics and Physics, Department of
  Macromolecular Physics, V Hole\v{s}ovi\v{c}k\'{a}ch 2, CZ-18000
  Praha 8, Czech Republic}

\author{Philipp Maass} \email{maass@uos.de}
\affiliation{Universit\"{a}t Osnabr\"{u}ck, Fachbereich
  Mathematik/Informatik/Physik, Barbarastra{\ss}e 7, D-49076
  Osnabr\"uck, Germany}

\date{December 19, 2024}

\begin{abstract}
In driven nonlinear systems, phase locking is an intriguing effect
leading to robust stationary states that are stable over extended
ranges of control parameters.  Recent experiments allow for exploring
microscopic mechanisms underlying such phenomena in collective
dynamics of micro- and nanoparticles. Here we show that phase-locked
dynamics of hardcore-interacting microparticles in a densely populated
periodic potential under time-periodic driving arises from running
solitary cluster waves.  We explain how values of phase-locked
currents are related to soliton velocities and why collective particle
dynamics synchronize with the driving for certain particle diameters
only.  Our analysis is based on an effective potential for the
solitary wave propagation and a unit displacement law, saying that the
total average shift of all particle positions per soliton period
equals one wavelength of the periodic potential.
\end{abstract}

\maketitle

Phase locking is a fundamental phenomenon in periodically driven
systems with nonlinear dynamics \cite{Pikovsky/etal:2001}.  A
prominent example are steps in the current-voltage characteristics of
Josephson junctions driven by radio-frequency fields, commonly known
as Shapiro steps \cite{Josephson:1962, Shapiro:1963}.  Generally,
these steps emerge when a system settles into a robust stationary
limit cycle, which is stable against changes of one or more of its
control parameters.  Analogous phase-locking effects were found in a
large variety of nonlinear systems like superconducting nanowires
\cite{Dinsmore/etal:2008, Bae/etal:2009, Ridderbos/etal:2019}, charge
density waves \cite{Gruener/Zettl:1985, Gruener:1988,
  Thorne/etal:1988, Hundley/Zettl:1989, Sinchenko/Monceau:2013},
skyrmions \cite{Reichardt/Reichardt:2015, Reichhardt/Reichhardt:2017a,
  Sato/etal:2020, Vizarim/etal:2020, Souza/etal:2024}, vortex lattices
\cite{Reichardt/etal:1999, Reichhardt/Nori:1999, Kolton/etal:2001}, as
well as in colloidal systems for both single-particle
\cite{Juniper/etal:2015, Juniper/etal:2017} and collective dynamics
\cite{Tierno/Fischer:2014, Juniper/etal:2015, Tierno/etal:2014,
  Bohlein/etal:2012, Bohlein/Bechinger:2012, Brazda/etal:2017,
  Brazda/etal:2018, Vanossi/etal:2020}.

For single-particle dynamics in a periodic potential $U(x)$ with
wavelength $\lambda$ under time-periodic driving $F(t)$ with period
$\tau$, phase-locked stationary states are synchronized with the
driving, i.e.\ have a period equal to an integer multiple $q\tau$ of
$\tau$, $q\in\mathbb{N}$.  For the particle to move between equivalent
points of $U(x)$ in the period $q\tau$, the distance between those
points must be an integer multiple $p\lambda$ of $\lambda$,
$p\in\mathbb{N}$.  In such a state, the particle moves with mean
velocity $v$ having the phase-locked value
\begin{equation}
v_{p,q}=\frac{p}{q}\frac{\lambda}{\tau}\,.
\label{eq:vpq}
\end{equation}
The stationary state is often robust against small changes of control
parameters, causing a mode with velocity $v_{p,q}$ to appear over a
certain range of e.g.\ amplitudes or frequencies of the time-periodic
driving.  Which $p/q$ occur depends on the type of microscopic
dynamics, driving, and shape of the periodic potential.  For
overdamped motion in a sinusoidal potential, in particular, only
integer multiples of $\lambda/\tau$ are robust phase-locked velocities
\cite{Waldram/Wu:1982}.

For many-particle dynamics, phase locking is less well understood and
often puzzling.  Here we explain how it arises in collective dynamics
of microparticles due to motion of solitary cluster waves. These waves
enable particle transport over potential barriers even if the barriers
are orders of magnitudes higher than the thermal energy $k_{\rm B}T$
\cite{Antonov/etal:2022a, Cereceda-Lopez/etal:2023}.

\begin{figure}[b!]
\includegraphics[width=0.7\columnwidth]{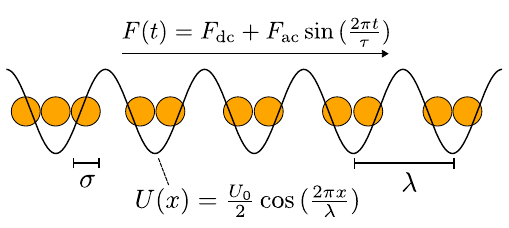}
\caption{Microparticles with hard-sphere diameter $\sigma$ moving in a
  sinusoidal potential $U(x)$ with wavelength $\lambda$ and potential
  barrier $U_0$. The particles perform an overdamped Brownian motion
  in a fluidic environment with temperature $T$ and are driven by a
  time-periodic force $F(t)$. We use $\lambda$, $U_0$ and
  $\lambda^2/\mu U_0$ as units of length, energy and time, where $\mu$
  is the particle mobility.}
\label{fig:model}
\end{figure}

Specifically, we consider overdamped Brownian motion of $N$
microparticles with hardcore interactions in $L/\lambda$ periods of a
densely populated sinusoidal potential $U(x)=(U_0/2)\cos(2\pi
x/\lambda)$ with barriers $U_0\gg k_{\rm B}T$ subject to periodic
boundary conditions.  The particles are driven by a time-periodic
force $F(t)=F_{\rm dc}+F_{\rm ac}\sin(2\pi t/\tau)$, see the
illustration in Fig.~\ref{fig:model}.
Their collective motion is explored by Brownian dynamics
  simulations.  Underlying Langevin equations are detailed in
Supplemental material (SM), including methods how to solve them
numerically and to determine relevant quantities, see below.

\begin{figure*}[t!]
\includegraphics[width=\textwidth]{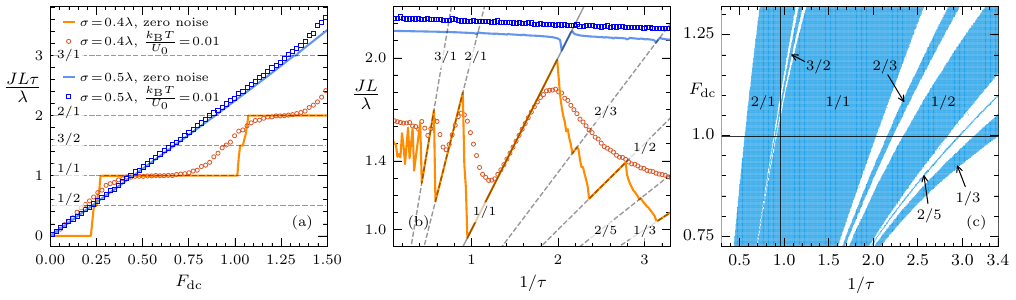}
\caption{Scaled stationary particle currents $J$ for the system shown
  in Fig.~\ref{fig:model} reflecting phase-locked dynamics. In (a),
  simulated currents are plotted as a function of the
  constant drag force $F_{\rm dc}$ at fixed $F_{\rm ac}=1.5$ and
  $1/\tau=3/\pi$. They show Shapiro steps at phase-locked values
  $J_{p,q}$ [Eq.~\eqref{eq:Jpq}].  In (b), simulated
  currents are plotted as a function of the driving frequency $1/\tau$
  at fixed $F_{\rm dc}=1$ and $F_{\rm ac}=1.5$ and the phase locked
  values $J_{p,q}$ show up as constant slopes in different intervals
  of $1/\tau$.  Data are shown for a system with $L/\lambda=20$
  potential wells and $N=21$ particles for two hard-sphere diameters
  $\sigma=0.4\lambda$ and $\sigma=0.5\lambda$ in the zero-noise limit
  ($k_{\rm B}T/U_0\to0$) and for a finite temperature, $k_{\rm
    B}T/U_0=0.01$ [legend in (a) applies also to panel (b)].  The
  currents in (a) and (b) are scaled with $L\tau/\lambda$ and
  $L/\lambda$, by which steps in (a) and slopes in (b) have rational
  values $p/q$, see Eq.~\eqref{eq:Jpq}.  (c) Diagram of
    phase locked currents $J_{p,q}$ for $\sigma=0.4\lambda$ in
    dependence of $1/\tau$ and $F_{\rm dc}$ for fixed $F_{\rm
      ac}=1.5$. The diagram was calculated from
      Eqs.~(S18), (S19) in SM.  Following the vertical and horizontal
    lines in the diagram yields the modes $J_{p,q}$ obtained in (a)
    and (b).}
\label{fig:Shapiro-steps}
\end{figure*}

Collective particle dynamics in the stationary state of this system
shows phase locking behavior with remarkable features, as reflected in
time-averaged currents shown in Fig.~\ref{fig:Shapiro-steps}:
\begin{list}{}{\setlength{\leftmargin}{1.8em}\setlength{\rightmargin}{0em}
\setlength{\itemsep}{0ex}\setlength{\topsep}{0ex}}

\item[(i)\,] Although we are confronted with complex many-particle
  dynamics, the phase-locked currents $J_{p,q}$ have values as if a
  single particle would be driven by a time-periodic force in a
  periodic potential with $L/\lambda$ periods of wavelength $\lambda$,
  i.e.\ with $\rho=1/L$ and $v_{p,q}=p\lambda/q\tau$:
\begin{equation}
J_{p,q}=\rho\, v_{p,q}=\frac{p}{q}\,\frac{\lambda}{L\tau}\,.
\label{eq:Jpq}
\end{equation}

\item[(ii)\;] Despite the potential being sinusoidal, fractional
  $J_{p,q}$ occur, with $p$ not divisible by $q$ in
  Eq.~\eqref{eq:Jpq}.

\item[(iii)] Occurrence of phase locking depends sensitively on the
  particle diameter $\sigma$: for certain $\sigma$, pronounced Shapiro
  steps appear ($\sigma=0.4\lambda$ in Fig.~\ref{fig:Shapiro-steps}),
  while they are nearly absent for other $\sigma$ ($\sigma=0.5\lambda$
  in Fig.~\ref{fig:Shapiro-steps}).

\end{list}
Ideal sharp Shapiro steps are obtained in the limit of zero noise
($k_{\rm B}T/U_0\to0$).  Pronounced ones can be clearly visible also
for temperatures well accessible in experiments
\cite{Cereceda-Lopez/etal:2021, Cereceda-Lopez/etal:2023}, see the
curves for $k_{\rm B}T/U_0=10^{-2}$ in Fig.~\ref{fig:Shapiro-steps}.

\begin{figure}[b!]
\includegraphics[width=\columnwidth]{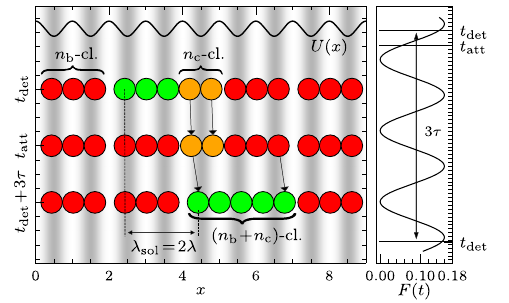}
\caption{Simulated particle configurations in a
  phase-locked steady state at different times for a soliton-carrying
  section of a system with $L/\lambda=23$ potential wells and $N=35$
  particles of size $\sigma=0.59\lambda$.  White and grey shaded
  stripes represent regions close to maxima and minima 
  of the sinusoidal potential $U(x)$ (solid line at top of
  graph), respectively.  Parameters of the time-periodic forcing $F(t)$, shown
  on the right side, are $F_{\rm dc}=F_{\rm ac}=0.08$ and
  $\tau=0.875$.  For meaning of labelings (time, colored particle
  clusters, distances), see main text.}
\label{fig:particle-configurations}
\end{figure}

We show that the puzzling features (i)-(iii) in the many-particle
dynamics originate from an effective potential for solitary wave
propagation and a law for the sum of all particle displacements
in one wave period.
 
Solitary cluster waves, in short cluster solitons, were
  recently predicted for overdamped Brownian motion under a constant
drag force \cite{Antonov/etal:2022a} and shortly after confirmed in
experiments \cite{Cereceda-Lopez/etal:2023}.  They characterize
running states when wells of a periodic potential are filled by
particles beyond a limit, where a mechanically stable state ceases to
exist.
 
The solitons turn out to mediate particle transport also under
additional time-periodic driving.
Figure~\ref{fig:particle-configurations} illustrates a propagation in
the zero-noise limit, where clusters of particles in contact form.
Most of the clusters in Fig.~\ref{fig:particle-configurations} consist
of three particles (red). We call them basic stable clusters, because
in the absence of time-periodic driving ($F_{\rm ac}=0$), they build
up a stable configuration with largest particle number
\cite{Antonov/etal:2024}.  For $F_{\rm ac}>0$, these basic clusters
of size $n_{\rm b}=3$ perform localized oscillatory motions.

In addition, we see a 2-cluster (orange), which at a time $t_{\rm
  det}$ is formed by detachment from a 3-cluster (green). Shortly
before the detachment, the 3- and 2-clusters formed a 5-cluster. After
a slight movement of the 2-cluster, it attaches to a 3-cluster at time
$t_{\rm att}$. The resulting 5-cluster moves over a long time interval
until time $t_{\rm det}+3\tau$. The configuration at this time instant
corresponds to that in the first row, at time $t_{\rm det}$, i.e.\ at
an infinitesimal time later, a 2-cluster detaches and the whole
process repeats itself. The sequential creation of slightly moving 2-
and 5-clusters propagates as a solitary wave. Its propagation is
demonstrated in the video of SM.

\begin{figure}[t!]
\includegraphics[width=\columnwidth]{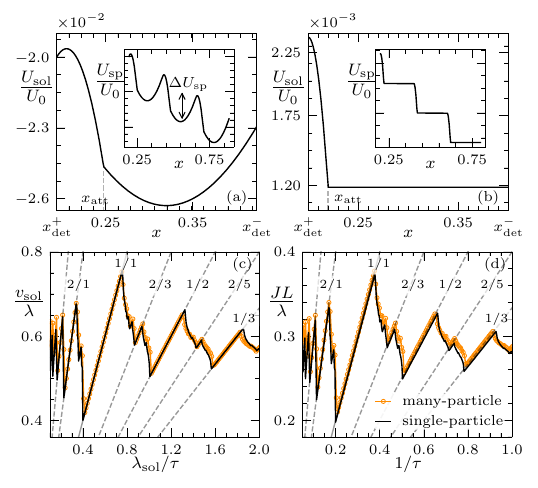}
\caption{(a) Soliton potential $U_{\rm sol}(x)$ [Eq.~\eqref{eq:Usol}]
  for parameters as in Fig.~\ref{fig:particle-configurations}
  ($\sigma=0.59\lambda$). It is shown for one period $\lambda_{\rm
    sp}=x_{\rm det}^--x_{\rm det}^+$ corresponding to one period
  $\lambda_{\rm sol}$ of the soliton motion.  At $x_{\rm att}$ it has
  a cusp because the propagating soliton cluster changes from the core
  to the composite cluster.  The inset shows three periods of the
  single-particle potential $U_{\rm sp}(x)$ with barrier $\Delta
  U_{\rm sp}$, obtained by stitching copies of $U_{\rm sol}(x)$.  (b)
  Soliton potential $U_{\rm sol}(x)$ and barrier-free single-particle
  potential $U_{\rm sp}(x)$ (inset) for $\sigma=0.6\lambda$.  (c), (d)
  Mean soliton velocity $v_{\rm sol}$ (symbols) as a function of
  $\lambda_{\rm sol}/\tau$ and mean particle current $J$ (multiplied
  by $L/\lambda$) as a function of $\lambda/\tau=1/\tau$ for
  $\sigma=0.59\lambda$. The simulated data show many
  linear segments with indicated rational slopes $p/q$.  Solid black
  lines were calculated from trajectories of a single particle moving
  in $U_{\rm sp}(x)$ and driven by $F(t)$.  In (c), single-particle
  velocities were scaled by $\lambda_{\rm sol}/\lambda_{\rm sp}$, and
  in (d) currents were obtained by weighting of instantaneous
  velocities and time-averaging described in the main text.}
\label{fig:Usol-v-J}
\end{figure}

In the example above, the 2-cluster represents the core soliton
cluster of size $n_{\rm c}=2$ and the 5-cluster the composite soliton
cluster of size $n_{\rm b}+n_{\rm c}=5$. We define the instantaneous
position of the soliton by the position of the leftmost particle in a
soliton cluster, i.e.\ the core or composite cluster. The soliton
motion is periodic in space with period $\lambda_{\rm sol}$. Phase
locking between soliton propagation and oscillatory force manifests
itself in that the soliton moves $p$ periods $\lambda_{\rm sol}$ in
$q$ time periods $\tau$ of the oscillatory force, implying
phase-locked values $(p/q)\lambda_{\rm sol}/\tau$ for the mean soliton
velocity $v_{\rm sol}$.  This fact is demonstrated in
Fig.~\ref{fig:Usol-v-J}(c) (orange line), where $v_{\rm sol}$ is shown
as a function of $\lambda_{\rm sol}/\tau$ for the parameters in
Fig.~\ref{fig:particle-configurations} ($p=1$, $\lambda_{\rm
  sol}=2\lambda$, and $q=3$).
  
The relation between solitary wave and particle dynamics enables us to
understand the phase locking in the many-particle system by
considering an effective local soliton potential $U_{\rm sol}(x)$, the
wavelength $\lambda_{\rm sol}$ of the solitary wave, and a unit
particle displacement law per soliton period.
 
The soliton potential gives the time-averaged mean force acting on the
core and composite soliton cluster ($F_{\rm ac}=0$).  It is
\begin{equation}
U_{\rm sol}(x)=\left\{\begin{array}{cl} U_{n_{\rm c}}(x)+C\,, &
\mbox{core cluster at $x$}\,,\\[0.5ex] U_{n_{\rm b}+n_{\rm c}}(x)\,, &
\mbox{composite cluster at $x$}\,,
\end{array}\right.
\label{eq:Usol}
\end{equation}
where $U_n(x)=\sum_{i=0}^{n-1} U(x+i\sigma)/n$. The constant
$C=U_{n_{\rm b}+n_{\rm c}}(x_{\rm att})-U_{n_{\rm c}}(x_{\rm att})$ is
added to make $U_{\rm sol}(x)$ continuous at $x=x_{\rm att}$.  The
range of positions $x$ in Eq.~\eqref{eq:Usol} covers the interval
$]x_{\rm det}^+,x_{\rm det}^-[$, where $x_{\rm det}^+$ is the soliton
    position right after the detachment of the core $n_{\rm
      c}$-cluster from the composite $(n_{\rm b}+n_{\rm c})$-cluster.
    The distance $x_{\rm det}^--x_{\rm det}^+$ is smaller than
    $\lambda$ and equals the empty gap between two successive $n_{\rm
      b}$ clusters. For the parameters in
    Fig.~\ref{fig:particle-configurations}, $U_{\rm sol}(x)$ is
    displayed in Fig.~\ref{fig:Usol-v-J}(a).

Given $U_{\rm sol}(x)$, one can map the soliton dynamics
onto that of a single quasiparticle moving in a potential $U_{\rm
  sp}(x)$ obtained by stitching together copies of $U_{\rm
  sol}(x)$, see the inset of Fig.~\ref{fig:Usol-v-J}(a). Remarkably,
$U_{\rm sp}(x)$ has an inherent tilt, since $U_{\rm sol}(x_{\rm
  det}^+)$ and $U_{\rm sol}(x_{\rm det}^-)$ are different for $F_{\rm
  dc}>0$. Without the tilt, the single-particle potential is periodic
with wavelength $\lambda_{\rm sp}=x_{\rm det}^--x_{\rm det}^+$.

Based on this periodic potential, we can understand the occurrence of
fractional Shapiro steps and the strong sensitivity of the phase
locking on the particle size. Fractional Shapiro steps can occur,
because the periodic potential is not sinusoidal.  The strong
sensitivity arises since even a small change of $\sigma$ can alter the
shape of $U_{\rm sp}(x)$ from one exhibiting barriers $\Delta U_{\rm
  sp}$ to one without barriers, and barriers are necessary to obtain
perceptible phase locking \cite{Kautz:1996}.  For example, $U_{\rm
  sp}(x)$ in Fig.~\ref{fig:Usol-v-J}(a) for $\sigma=0.59\lambda$ has
barriers, while it has no barriers for $\sigma=0.6\lambda$, see insets
of Figs.~\ref{fig:Usol-v-J}(a), (b).  Likewise, for
$\sigma=0.4\lambda$ in Fig.~\ref{fig:Shapiro-steps}, $U_{\rm sp}(x)$
has barriers, and no ones for $\sigma=0.5\lambda$, see SM.

By generating trajectories of single-particle motion in the force
field $F(t)-\dd U_{\rm sp}/\dd x$ as described in SM, we can moreover
obtain $v_{\rm sol}$ and $J$ in the many-particle system with high
accuracy after following transformations.

The velocity $v_{\rm sol}$ is obtained by rescaling the mean
single-particle velocity $v_{\rm sp}$. In one period $\lambda_{\rm
  sp}$ of the single-particle motion, the associated soliton moves a
distance $\lambda_{\rm sol}>\lambda_{\rm sp}$.  This is because one
needs to take into account the jumps $(\lambda_{\rm sol}-\lambda_{\rm
  sp})$ occurring in soliton trajectories at detachment events. Hence,
$v_{\rm sol} = (\lambda_{\rm sol}/\lambda_{\rm sp})v_{\rm sp}$.

For calculating $\lambda_{\rm sol}$, we consider two successive points
where a core cluster attaches to a basic cluster.  The distance
between the first particles of the two basic clusters in the
attachment events have distance $\lceil n_{\rm
  b}(\sigma/\lambda)\rceil\lambda$, where $\lceil x\rceil$ is the
smallest integer larger than $x$.  Hence,
\begin{equation}
\lambda_{\rm sol}=\left\lceil n_{\rm
  b}\,\frac{\sigma}{\lambda}\right\rceil\lambda\,.
\end{equation}
For $n_{\rm b}=3$ and $\sigma=0.59\lambda$ in
Fig.~\ref{fig:particle-configurations}, $\lceil n_{\rm
  b}(\sigma/\lambda)\rceil=2$, see the spacing between the dotted
lines.  Rescaled single-particle velocities are in excellent agreement
with $v_{\rm sol}$ in the many-particle system, see the black solid
line in Fig.~\ref{fig:Usol-v-J}(c).

From the phase-locked soliton velocities $v_{p,q}=(p/q)\lambda_{\rm
  sol}/\tau$ and the single-particle picture one may expect that
$J_{p,q}=v_{p,q}/L$ gives the phase-locked many-particle currents.
However, the current in Fig.~\ref{fig:Usol-v-J}(d) does not display
values $LJ_{p,q}/\lambda=(p/q)2/\tau$ for $\lambda_{\rm
  sol}=2\lambda$, but we obtain by a factor 1/2 smaller values
$LJ_{p,q}/\lambda=(p/q)/\tau$, i.e.\ $J_{p,q}=v_{p,q}/2L$.
Thus, the naive picture of a soliton
as a current carrying quasiparticle does not apply.
 
To obtain the currents from single-particle trajectories, one needs to
include a weighting factor $n(x)$ accounting for the fact that $n_{\rm
  c}$ (or $n_{\rm b}+n_{\rm c}$) particles are moving in the
many-particle system if the single particle is at a position $x$ of a
core (or composite) soliton cluster.  Then, $J$ is obtained by
time-averaging $n(x_{\rm sp}(t))v_{\rm sp}(t)/L$, where $x_{\rm
  sp}(t)$ and $v_{\rm sp}(t)$ are the position and velocity of the
single particle at time $t$.  The so-calculated currents, shown as
black solid line in Fig.~\ref{fig:Usol-v-J}(d), reproduce very well
the many-particle currents.

The weighting of single-particle velocities gives insight why
$J_{p,q}$ differs from $v_{p,q}/L$, but it does not explain why
$J_{p,q}=v_{p,q}/2L$.
 This relation follows from the fact that the
sum of all $N$ particle displacements is $p\lambda$ if the soliton moves $p$
wavelengths $\lambda_{\rm sol}$, i.e.\ a unit displacement law holds on
average:
\begin{equation}
\frac{1}{p}\left[\sum_{i=1}^N \Delta
  x_i\right]_{\parbox[b]{3.5em}{\small $p$
    soliton\\ periods}}=\lambda\,.
\label{eq:displacement-law}
\end{equation}
Accordingly, for the $p/q$-phase locked soliton motion, the
phase-locked values of the particle current is
$J_{p,q}=(1/L)\sum_{i=1}^N \Delta
x_i/q\tau=(p/q)\lambda/L\tau=v_{p,q}/2L$. That is, the
  factor 1/2 follows from the fact that the soliton moves
  $\lambda_{\rm sol}$ in one period of the driving, while the mean
  total displacement of all particles per soliton period is
  $\lambda=\lambda_{\rm sol}/2$.

Using $U_{\rm sp}(x)$, we can also predict the intervals of driving
parameters where phase-locked modes with certain $p$ and $q$ occur in
the limit of zero noise.  The procedure is described in SM and
demonstrated by a diagram of phase-locked modes in
Fig.~\ref{fig:Shapiro-steps}(c). When following the horizontal and
  vertical lines in the diagram, we can infer the phase-locked current
  modes shown in Figs.~\ref{fig:Shapiro-steps}(a),(b).  For example,
  the modes with $p/q=2/1$, $1/1$, and $1/2$ appear
  for $1/\tau$ in the intervals 0.56-0.87, 0.92-2.03, and 2.33-2.80 in
  Fig.~\ref{fig:Shapiro-steps}(c), which is in excellent agreement
  with the corresponding intervals of $1/\tau$ in
  Fig.~\ref{fig:Shapiro-steps}(b).

Let us now discuss the impact of thermal noise on the phase locking
that is rather strong in general \cite{Tekic/Hu:2008, Mali/etal:2012,
  Mali/etal:2020}.  Its disturbing influence is the smaller the larger
$\Delta U_{\rm sp}/k_{\rm B}T$. Studies of Josephson junctions show
that phase locking can be destroyed for thermal energies much less
than the barrier of the periodic potential \cite{Kautz:1996}.
For the system in Fig.~\ref{fig:Shapiro-steps} we
calculated $\Delta U_{\rm sp}/U_0$ as a function of $\sigma$, see SM.
The $\Delta U_{\rm sp}$ are much smaller than $U_0$ and run through
maxima between zero points occurring at diameters
$\sigma=(n-1)/n\lambda$, $n=2,3,\ldots$
\cite{Antonov/etal:2022a}. Overall, the barriers $\Delta U_{\rm
  sp}/k_{\rm B}T$ decrease with $\sigma$, making small diameters
favorable for studies of phase-locking in the many-particle system at
finite thermal noise. For $\sigma=0.4\lambda$, barriers $\Delta U_{\rm
  sp}$ are about $0.18U_0$ and the phase locking is well visible for
$k_{\rm B}T=0.01U_0$.

In summary, we have shown how phase-locked dynamics arise for
microparticles driven by a time-periodic force with frequency $1/\tau$
through a densely populated sinusoidal potential in a fluidic
environment.  Responsible are phase-locked dynamics of solitary
cluster waves with a wavelength $\lambda_{\rm sol}$ that is an integer
multiple of the wavelength $\lambda$ of the sinusoidal potential.  To
describe the wave propagation, we have constructed an effective
periodic potential for the time-averaged soliton motion. This
potential is nonsinusoidal, leading to fractional phase-locked values
$(p/q)\lambda_{\rm sol}/\tau$ of the mean soliton velocity
($p,q\in\mathbb{N}$). It allows one to determine diagrams of phase-locked modes
and to estimate temperature limits, where phase-locking in the many-particle system
becomes destroyed by thermal noise.

The solitary cluster waves mediate particle displacements, leading to
fractional phase-locked current values in the many-particle system.  A
total sum of particle displacements $p\lambda$ is generated if a
soliton travels $p$ periods $\lambda_{\rm sol}$.  This corresponds to
a remarkable unit displacement law: per period $\lambda_{\rm sol}$ on
average, the sum of particle displacements is $\lambda$. As a
consequence, particle currents in the many-particle systems exhibit
phase-locked values $(p/q)\lambda/\tau$ equal to those of a single
driven particle in the periodic potential.

The phase-locking is well observable in the presence of thermal noise,
and Shapiro steps can remain clearly visible at ambient temperatures
in experiments with colloidal particles \cite{Juniper/etal:2015,
  Juniper/etal:2016, Juniper/etal:2017, Cereceda-Lopez/etal:2023}.
In these experiments, possible imperfections need to be
  considered also, as, e.g., a weak amplitude modulation of the
  periodic optical potential and small polydispersity.  We have checked
  that the phase locking effects shown in
  Figs.~\ref{fig:Shapiro-steps}(a),(b) are robust against such
  imperfections.

For larger system sizes or larger particle numbers, it is possible
that several solitons are propagating \cite{Cereceda-Lopez/etal:2023,
  Antonov/etal:2024}. In such cases, we conjecture that the unit
displacement law applies per soliton, giving phase-locked currents
$J_{p,q}=(p/q)N_{\rm sol}\lambda/\tau$, where $N_{\rm sol}$ is the
number of solitons.

Analogous synchronized solitary wave propagation are
  expected to occur for other periodic potentials and repulsive
  particle interactions.  An open question is what types of mechanisms
  lead to phase-locking of many-particle currents in underdamped
  Brownian dynamics.

Our theoretical approach can be useful for interpreting
experiments. This includes already existing observations, as the
phase-locked kink dynamics in a chain of magnetically interacting
particles \cite{Juniper/etal:2015}.  It was associated with a density
wave, which resembles the solitary cluster wave considered here.
Extended regimes of constant phase-locked particle currents as a
function of the driving amplitude have been observed in driven motion
of superparamagnetic particles across a magnetic bubble lattice
\cite{Tierno/Fischer:2014, Tierno/etal:2014}, and in molecular
dynamics simulations \cite{Paronozzi-Ticco/etal:2016} modelling
experiments on particle transport across colloidal monolayers
\cite{Bohlein/etal:2012, Bohlein/Bechinger:2012, Brazda/etal:2018,
  Vanossi/etal:2020}.  Detailed information on individual particle
motions accessible in such experiments can be utilized to test and
extend the concepts of an effective potential for cluster solitons and
the unit displacement law for connecting solitary wave motion to
many-particle transport.

\begin{acknowledgments}
We thank A.~Antonov and S.~Schweers for their work on code
development, and P.~Tierno for very helpful discussions on soliton
dynamics in experiments.  We gratefully acknowledge financial support
by the Czech Science Foundation (Project No.\ 23-09074L) and the
Deutsche Forschungsgemeinschaft (Project No.\ 521001072), and the use
of a high-performance computing cluster funded by the Deutsche
Forschungsgemeinschaft (Project No.\ 456666331).
\end{acknowledgments}
 

%

\onecolumngrid
\newpage
\renewcommand{\theequation}{S\arabic{equation}}
\renewcommand{\thefigure}{S\arabic{figure}}
\setcounter{equation}{0}
\setcounter{figure}{0}

\begin{center}
\setcounter{page}{1}
{\large\bf Supplemental Material for}\\[2ex]
{\large\bf Phase locking and fractional Shapiro steps in collective dynamics of microparticles}\\[2ex]
Seemant Mishra,$^1$ Artem Ryabov,$^{2}$ and Philipp Maass$^1$\\[1ex]
$^1$\textit{Universit\"{a}t Osnabr\"{u}ck, Fachbereich Mathematik/Informatik/Physik,\\
Barbarastra{\ss}e 7, D-49076 Osnabr\"uck, Germany}\\[1ex]
$^2$\textit{Charles
  University, Faculty of Mathematics and Physics, Department of Macromolecular Physics,\\ 
  V Hole\v{s}ovi\v{c}k\'{a}ch 2, CZ-18000 Praha 8, Czech Republic}\\[1ex]
\end{center}

\setstretch{1.5}
\vspace{1ex}\noindent

In Sec.~\ref{sec:langevin-equations} of this Supplemental Material, we
discuss Langevin equations for the particle motion and the
methods used to solve them numerically.
Section~\ref{sec:soliton-potential} gives a detailed derivation of the
soliton potential and Sec.~\ref{sec:sp-dynamics} describes the
effective single-particle dynamics reflecting  the soliton motion.  
The particle size dependence of barriers $\Delta U_{\rm sp}$ for
solitary wave propagation is exemplified in Sec.~\ref{sec:DeltaUsp}
for the parameters in Fig.~2 of the main text.  In
Sec.~\ref{sec:mode-diagram}, we explain how to calculate diagrams of
phase-locked current modes $J_{p,q}$.

\section{Langevin equations and
methods of their numerical solution}
\label{sec:langevin-equations}
Overdamped Brownian motion of $N$ hard spheres of diameter $\sigma$ in
the sinusoidal potential
\begin{equation}
U(x)=\frac{U_0}{2}\cos(\frac{2\pi x}{\lambda})
\end{equation}
driven by the time-periodic force
\begin{equation}
F(t)=F_{\rm dc}+F_{\rm ac}\sin(\frac{2\pi t}{\tau})
\end{equation}
is described by the Langevin equations
\begin{equation}
\frac{\dd x_i}{\dd t}=\mu\left[F(t)-\frac{\dd U}{\dd
    x_i}\right]+\sqrt{2D}\,\eta_i(t)\,,\hspace{1em}i=1,\ldots,N\,,
\label{eq:langevin}
\end{equation}
where $\mu$ is the mobility, $D=k_{\rm B}T\mu$, and $T$ the
temperature.  The $\eta_i(t)$ are stationary Gaussian processes with
mean $\langle\eta_i(t)\rangle=0$ and correlation functions $\langle
\eta_i(t)\eta_j(t')\rangle=\delta_{ij}\delta(t-t')$.  Periodic
boundary conditions are applied for a system with $L/\lambda$
potential wells.  The hard-sphere interaction implies
\begin{equation}
|x_i-x_j|\ge\sigma\,.
\label{eq:hs-constraint}
\end{equation}

Results presented in the main text were obtained by numerical
integration of Langevin equations \eqref{eq:langevin} under
consideration of the hard-sphere constraint \eqref{eq:hs-constraint},
using the recently developed Brownian cluster dynamics algorithm
\cite{Antonov/etal:2022c, Antonov/etal:2025} with time step $\Delta
t=10^{-4}\lambda^2/\mu U_0$.  We checked these results against
Brownian dynamics simulations, where the hard-sphere interaction is
treated based on elastic collision schemes \cite{Strating:1999,
  Scala:2012}.

In our numerical treatment, we rescaled quantities to dimensionless
variables by taking $\lambda$, $U_0$, and $\lambda^2/\mu U_0$ as units
of length, energy, and time, respectively. All quantities were sampled
after a transient time, where the system was brought into a stationary
state.

Particle currents in the stationary states were obtained by taking
averages of the instantaneous values
\begin{equation}
J(t)=\frac{1}{L}\sum_{i=1}^N \frac{\dd x_i}{\dd t}\,.
\label{eq:J(t)}
\end{equation}
The instantaneous velocities $v_i(t)=\dd x_i(t)/\dd t$
in Eq.~\eqref{eq:J(t)} were calculated by taking 
difference quotients $[x_i(t+\Delta t)-x_i(t)]/\Delta t$.
The currents $J$ shown in Figs.~2(a), 2(b), and 4(c), 4(d)
are the time-averaged instantaneous currents $J(t)$ 
in the steady state, i.e.\ $J=\int_0^{M\tau} \dd t J(t)/M\tau$ with $M\gg1$.

\section{Soliton potential}
\label{sec:soliton-potential}
\begin{figure}[t!]
\includegraphics[width=0.5\textwidth]{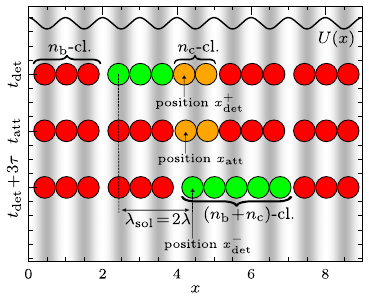}
\caption{Soliton propagation from Fig.~3 of the main text, where the
  positions $x_{\rm det}^+$, $x_{\rm att}$, and $x_{\rm det}^-$ of the
  soliton are indicated.  The position $x_{\rm det}^+$ is that of the
  core soliton $n_{\rm c}$-cluster right after its detachment from a
  composite cluster, $x_{\rm att}$ is the position, when the core
  $n_{\rm c}$-cluster attaches to an $n_{\rm b}$-cluster, and $x_{\rm
    det}^-$ is the position of the composite $(n_{\rm b}+n_{\rm
    c})$-cluster right before the detachment of a new core $n_{\rm
    c}$-cluster.}
\label{fig:soliton-propagation}
\end{figure}

For modeling the soliton dynamics in a single-particle description, we
determine the time-averaged mean forces acting on the clusters
involved in the wave propagation. These clusters
are the core soliton cluster of size $n_{\rm c}$ and the composite
soliton cluster of size $n_{\rm b}+n_{\rm c}$, see Figs.~3 and~\ref{fig:soliton-propagation}.

The time-averaged force acting on a single particle at position $x$ is
\begin{equation}
\bar F(x)=F_{\rm dc}-\frac{\dd U(x)}{\dd x}\,.
\label{eq:F(x)}
\end{equation}
In an $n$-cluster, particles keep in touch at positions $x_1(t)$,
$x_2(t)=x_1(t)+\sigma, \ldots$, $x_n(t)=x_1(t)+(n\!-\!1)\sigma$.  The
center of mass velocity of such $n$-cluster is given by the mean
external force
\begin{equation}
\bar F_n(x_1)=\frac{1}{n}\,[\bar F(x_1)+\ldots+\bar F(x_n)]
\end{equation}
acting on the $n$-cluster. We have defined the position of the
$n$-cluster to be the position $x_1$ of the leftmost particle in the
$n$-cluster.

The force $\bar F_n$ can be written as
\begin{equation}
\bar F_n(x)=F_{\rm dc}-\frac{\dd}{\dd x} U_n(x)\,,
\end{equation}
where $U_n(x)$ is the potential of an $n$-cluster given after Eq.~(3) in the main text:
\begin{equation}
U_n(x)=\frac{1}{n}\sum_{i=0}^{n-1} U(x+i\sigma)\,.
\end{equation}

As discussed in the main text and illustrated in Figs.~3 and~\ref{fig:soliton-propagation}, 
we can describe one period of soliton propagation as follows: Initially, a core soliton cluster
of size $n_{\rm c}$ detaches from a composite cluster at a position
$x_{\rm det}^+$. The plus in the superscript indicates that it is the position of the
core cluster (soliton) \textit{right after} its detachment.  Right
before the detachment, the soliton position is that of the composite
cluster, i.e.\ $x_{\rm det}^+-n_{\rm b}\sigma$.  In the time after the
detachment, the $n_{\rm c}$-cluster propagates until attaching to an
$n_{\rm b}$-cluster at position $x_{\rm att}$. Due to the attachment,
a composite cluster of size $n_{\rm b}+n_{\rm c}$ forms. The soliton
position at the moment of attachment is $x_{\rm att}$ and thereafter
it is the position of the moving composite cluster. The motion of the
newly formed composite cluster continues until it reaches position
$x_{\rm det}^-$. At this moment, the process starts anew, i.e.\ a core
$n_{\rm c}$-cluster detaches from the composite cluster at position
$x_{\rm det}^-+n_{\rm b}\sigma$. In each detachment event of a core
$n_{\rm c}$-cluster, the soliton position jumps by $n_{\rm b}\sigma$.

The soliton potential describes the continuous soliton motion in the
interval $[x_{\rm det}^+,x_{\rm det}^-]$, which is given by the
cluster potential $U_{n_{\rm c}}(x)$ in the subinterval $[x_{\rm
    det}^+,x_{\rm att}]$ and by the cluster potential $U_{n_{\rm
    b}+n_{\rm c}}(x)$ in the subinterval $[x_{\rm att},x_{\rm
    det}^-]$:
\begin{equation}
U_{\rm sol}(x)=\left\{\begin{array}{cl}
U_{n_{\rm c}}(x)+C\,, & x_{\rm det}^+\le x\le x_{\rm att}\,,\\[0.5ex] 
U_{n_{\rm b}+n_{\rm c}}(x)\,, & x_{\rm att}\le x\le x_{\rm det}^-\,.
\end{array}\right.
\label{eq:Usol-suppl}
\end{equation}
This is Eq.~(3) of the main text, where the constant $C=U_{n_{\rm
    b}+n_{\rm c}}(x_{\rm att})-U_{n_{\rm c}}(x_{\rm att})$ is added to
$U_{n_{\rm c}}(x)$ for making $U_{\rm sol}(x)$ continuous at $x_{\rm
  att}$.  The length $x_{\rm det}^--x_{\rm det}^+$ of the continuous
soliton motion gives the wavelength $\lambda_{\rm sp}$ of the
associated single-particle potential $U_{\rm sp}(x)$ in
Eq.~\eqref{eq:Usp} below.

While thermal noise is neglected in the construction of the soliton
potential, it is considered in the single-particle description of the
soliton dynamics, see Eq.~\eqref{eq:langevin-single} below.

\section{Single-particle description of soliton dynamics}
\label{sec:sp-dynamics}
The single-particle potential $U_{\rm sp}(x)$ obtained by stitching
copies of the soliton potential $U_{\rm sol}(x)$ in
Eq.~\eqref{eq:Usol-suppl} and Eq.~(3) of the main text is
\begin{equation}
U_{\rm sp}(x)=\sum_{j=-\infty}^\infty I_j(x)\,[U_{\rm
    sol}(x-j\lambda_{\rm sp})-j\Delta U_{\rm sol}]\,,
\label{eq:Usp}
\end{equation}
where $\lambda_{\rm sp}=x_{\rm det}^- -x_{\rm det}^+$, see
Fig.~\ref{fig:soliton-propagation},
\begin{equation}
\Delta U_{\rm sol}=\lim_{\varepsilon\to0^+}\left[U_{\rm sol}(x_{\rm
    det}^++\varepsilon)-U_{\rm sol}(x_{\rm
    det}^--\varepsilon)\right]\,,
\end{equation}
and $I_j(x)$ is the indicator function for the interval $[x_{\rm
    det}^++j\lambda_{\rm sp},x_{\rm det}^-+j\lambda_{\rm sp}[$,
\begin{equation}
I_j(x)=\left\{\begin{array}{cc} 1\,, & \mbox{for $x\in[x_{\rm
      det}^++j\lambda_{\rm sp},x_{\rm det}^-+j\lambda_{\rm
      sp}[$}\,,\\ 0\,, & \mbox{otherwise.}
\end{array}\right.
\end{equation}

The overdamped Brownian motion of a single particle in the potential
$U_{\rm sp}(x)$ driven by $F(t)$ is described by the Langevin equation
\begin{equation}
\frac{\dd x}{\dd t}=\mu\left[F(t)-\frac{\dd U_{\rm sp}}{\dd
    x}\right]+\sqrt{2D(x)}\,\eta(t)\,,
\label{eq:langevin-single}
\end{equation}
where $\eta(t)$ is a stationary Gaussian process with
$\langle\eta(t)\rangle=0$,
$\langle\eta(t)\eta(t')\rangle=\delta(t-t')$, and $D(x)$ is a
space-dependent diffusion coefficient
\begin{equation}
D(x)=\frac{D}{n(x)}\,,
\end{equation}
with
\begin{equation}
n(x)=\left\{\begin{array}{cc} n_{\rm c}\,, & \mbox{for $x\in[x_{\rm
      det}^++j\lambda_{\rm sp},x_{\rm att}+j\lambda_{\rm sp}[$,
      $j\in\mathbb{Z}$}\,,\\ n_{\rm b}+n_{\rm c}\,, &
    \mbox{otherwise.}
\end{array}\right.
\end{equation}
The Langevin equation \eqref{eq:langevin-single} is in Ito
interpretation.

\section{Dependence of barriers $\Delta U_{\rm sp}$ on particle size}
\label{sec:DeltaUsp}

\begin{figure}[t!]
\includegraphics[width=0.7\textwidth]{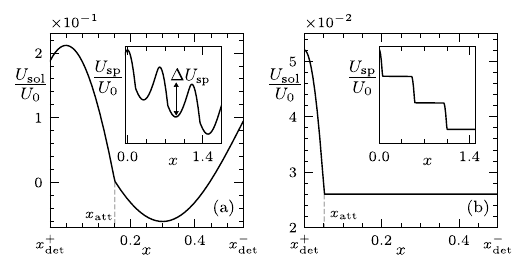}
\caption{Single-particle potentials [Eq.~\eqref{eq:Usp}]
at $F_{\rm dc}=1$ for (a)
  $\sigma=0.4\lambda$ and (b) $\sigma=0.5\lambda$.}
\label{fig:Usp_for_Fig2}
\end{figure}

Single-particle potentials $U_{\rm sp}(x)$ at $F_{\rm dc}=1$ are
displayed in Fig.~\ref{fig:Usp_for_Fig2} for (a) $\sigma=0.4\lambda$
and (b) $\sigma=0.5\lambda$. For $\sigma=0.4\lambda$, $U_{\rm sp}(x)$
has pronounced barriers $\Delta U_{\rm sp}$, giving rise to the robust
phase-locking in Fig.~2 of the main text, which remains largely
observable in the presence of thermal noise.

\begin{figure}[t!]
\includegraphics[width=0.6\textwidth]{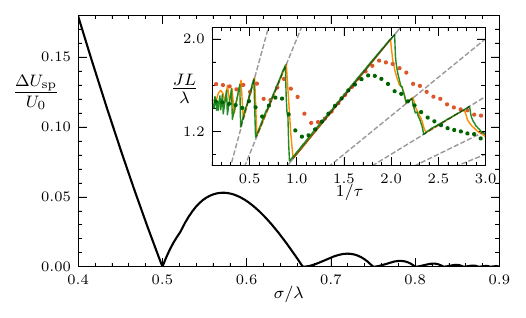}
\caption{Variation of barriers $\Delta U_{\rm sp}$ of the
  single-particle potential with the particle diameter $\sigma$ for
  the system in Fig.~2 of the main text.  The inset shows the
  many-particle currents from Fig.~2(b) ($\sigma=0.4\lambda$, orange)
  in comparison with currents calculated from single-particle
  simulations (green) in the potential $U_{\rm sp}$. Lines
  mark the results in the zero-noise limit and symbols for thermal
  noise at $k_{\rm B}T/U_0=0.01$.}
\label{fig:Usp_barriers}
\end{figure}

How $\Delta U_{\rm sp}$ depends on the particle diameter $\sigma$ is
shown in Fig.~\ref{fig:Usp_barriers} for the system in Fig.~2 with
$L/\lambda=20$ potential wells and $N=21$ particles.  The $\Delta
U_{\rm sp}$ are much smaller than $U_0$ and run through maxima between
zero points occurring at magic diameters $\sigma=\lambda (n-1)/n$,
$n=2,3,\ldots$ \cite{Antonov/etal:2022a}. With increasing $\sigma$,
the maxima of $\Delta U_{\rm sp}$ decrease, making small diameters
favorable for studies of phase-locking in the presence of thermal
noise. For $\sigma=0.4\lambda$ in Fig.~2, $\Delta U_{\rm sp}$ is about
$0.18U_0$, which is sufficient to obtain phase-locked currents for
$k_{\rm B}T=0.01U_0$.

A comparison of the many-particle currents in Fig.~2(b) with those
obtained from single-particle simulations is shown in the inset of
Fig.~\ref{fig:Usp_barriers}. In the zero-noise limit, the
single-particle results agree well with the many-particle data, as in
Fig.~4(d) of the main text.  For $k_{\rm B}T=0.01U_0$, the widths of
frequency intervals, where phase-locked values $J_{p,q}$ manifest
themselves as slopes in the curves are nearly the same. Interestingly,
the currents obtained from single-particle modeling are slightly
shifted to smaller frequencies and smaller current values, an effect
related to the construction of $U_{\rm sp}(x)$ in the zero-noise
limit.

\section{Diagram of phase-locked modes}
\label{sec:mode-diagram}

Knowing $U_{\rm sp}(x)$ with wavelength $\lambda_{\rm sp}$, we can
predict the parameter intervals, where specific phase-locked modes
$p/q$ occur.  Starting point is the construction of the one-period
propagator $G(x;\Gamma)$ for the single-particle dynamics according to
Eq.~\eqref{eq:langevin-single} in the absence of noise
[$D(x)=0$]. This propagator depends on the set of driving parameters
$\Gamma=(F_{\rm dc}, F_{\rm ac}, \tau)$ and gives the final position
$x_{\rm f}=G(x;\Gamma)$ after one period $\tau$ of the driving, if the
particle starts at the initial position $x$. For brevity, we suppress
the argument $\Gamma$ in the propagator in the following.

For a phase-locked periodic mode with value $q$, the particle must be
at an equivalent position in the periodic potential after $q$ periods
of the driving, i.e.\ the difference between the final position
$x_{\rm f}^{(q)}$ after the $q$ periods and the initial position $x$
must be an integer multiple of $\lambda_{\rm sp}$.  The final position
$x_{\rm f}^{(q)}$ is given by the $q$-period propagator $G^{(q)}(x)$,
which is obtained by $q$-fold composition of the one-period
propagator:
\begin{equation}
x_{\rm f}^{(q)}(x)=G^{(q)}(x)=\underbrace{G\circ\ldots\circ G}_{\mbox{$q$ times}}(x)\,.
\label{eq:xfinal}
\end{equation}
The smallest $q$ for which there exists an $x_\ast\in[0,\lambda_{\rm
    sp}[$, where $x_{\rm f}^{(q)}(x_\ast)-x_\ast$ is an
    integer multiple of $\lambda_{\rm sp}$, is the $q$ of a
    phase-locked mode. The $q$ of such a mode satisfies
\begin{equation}
[G^{(q)}(x_\ast)\hspace{-0.7em}\mod \lambda_{\rm sp}]-x_\ast=0\,,
\label{eq:q-determ}
\end{equation}
where $[a\hspace{-0.5em}\mod b]\in[0,b[$ is the remainder when $a$ is
    divided by $b$. The number $p$ of wavelengths $\lambda_{\rm sp}$
    by which the particle is displaced after $q$ periods of the
    driving is
\begin{equation}
p=\frac{|G^{(q)}(x_\ast)-x_\ast|}{\lambda_{\rm sp}}\,.
\label{eq:p-determ}
\end{equation}
Equations~\eqref{eq:q-determ} and \eqref{eq:p-determ} determine $q$
and $p$ values of a phase locked mode.

A fixed point $x_\ast$ of $[G^{(q)}(x)\hspace{-0.4em}\mod \lambda_{\rm
    sp}]$ in Eq.~\eqref{eq:q-determ} corresponds to a limit cycle of
$[G^{(1)}(x)\hspace{-0.4em}\mod \lambda_{\rm sp}]$ that runs through
$q$ points.  Each of these points is a fixed point $x_\ast$ of
$[G^{(q)}(x)\hspace{-0.4em}\mod \lambda_{\rm sp}]$, i.e.\ if there is
one fixed point $x_\ast$ according to Eq.~\eqref{eq:q-determ}, there
will be $q\!-\!1$ further fixed points belonging to the same limit
cycle and satisfying Eq.~\eqref{eq:q-determ}.  A fixed point $x_\ast$
is stable against small perturbations if
$|\partial_xG^{(q)}(x_\ast)|<1$ \cite{Strogatz:2015}.  Unstable fixed
points are excluded from the analysis.  In the numerical procedure
described next for determining $p/q$, we moreover checked for 20 sets
of driving parameters $\Gamma$ that the attraction basin of the fixed
points covers the whole interval $x\in[0,\lambda_{\rm sp}[$: for each
    $\Gamma$, iteratively mapping of 100 randomly chosen initial
    positions $x\in[0,\lambda_{\rm sp}[$ always converged to same
        limit cycle.

In the numerical procedure, we search for zeros of
$H^{(q)}(x)=[G^{(q)}(x)\hspace{-0.4em}\mod \lambda_{\rm sp}]-x$ in
$[0,\lambda_{\rm sp}[$, i.e.\ fixed points of
    [$G^{(q)}(x)\hspace{-0.4em}\mod \lambda_{\rm sp}]$. Starting with
    $q=1$, $q$ is incremented by one until a root is found. This gives
    the $q$ of the phase-locked mode.

To determine the zeros of $H^{(q)}(x)$, we divide the interval
$[0,\lambda_{\rm sp}[$ into $M=50$ equidistant points $x_i$,
    $i=1,\ldots M$, and identify all successive points $x_i$ and
    $x_{i+1}$, where $H^{(q)}(x)$ changes sign.  There can be several
    of such pairs, i.e.\ we obtain a list of intervals
    $[x_i^{(\alpha)},x_{i+1}^{(\alpha)}]$, $\alpha=1,\ldots,m<M$.
    Applying the Van Wijngaarden–Dekker–Brent method
    \cite{Press/etal:2007} on each of these intervals, the points
    $x_0^{(\alpha)}$ of sign change in all intervals $\alpha$ are
    determined with high accuracy of 11 significant digits.  Not all
    $x_0^{(\alpha)}$ are roots of Eq.~\eqref{eq:q-determ}, because
    $H^{(q)}(x)$ can have a jump at $x_0^{(\alpha)}$,
    i.e.\ $|H^{(q)}(x_0^{(\alpha)}-0)-H^{(q)}(x_0^{(\alpha)}+0)|>0$. In
    the procedure, we chose
    $|H^{(q)}(x_0^{(\alpha)}-\delta)-H^{(q)}(x_0^{(\alpha)}+\delta)|>\epsilon$
    with $\delta=10^{-6}$ and $\epsilon=10^{-3}$ as criterion for
    excluding $x_0^{(\alpha)}$ as roots of Eq.~\eqref{eq:q-determ}.

\begin{figure}[t!]
\includegraphics[width=0.7\textwidth]{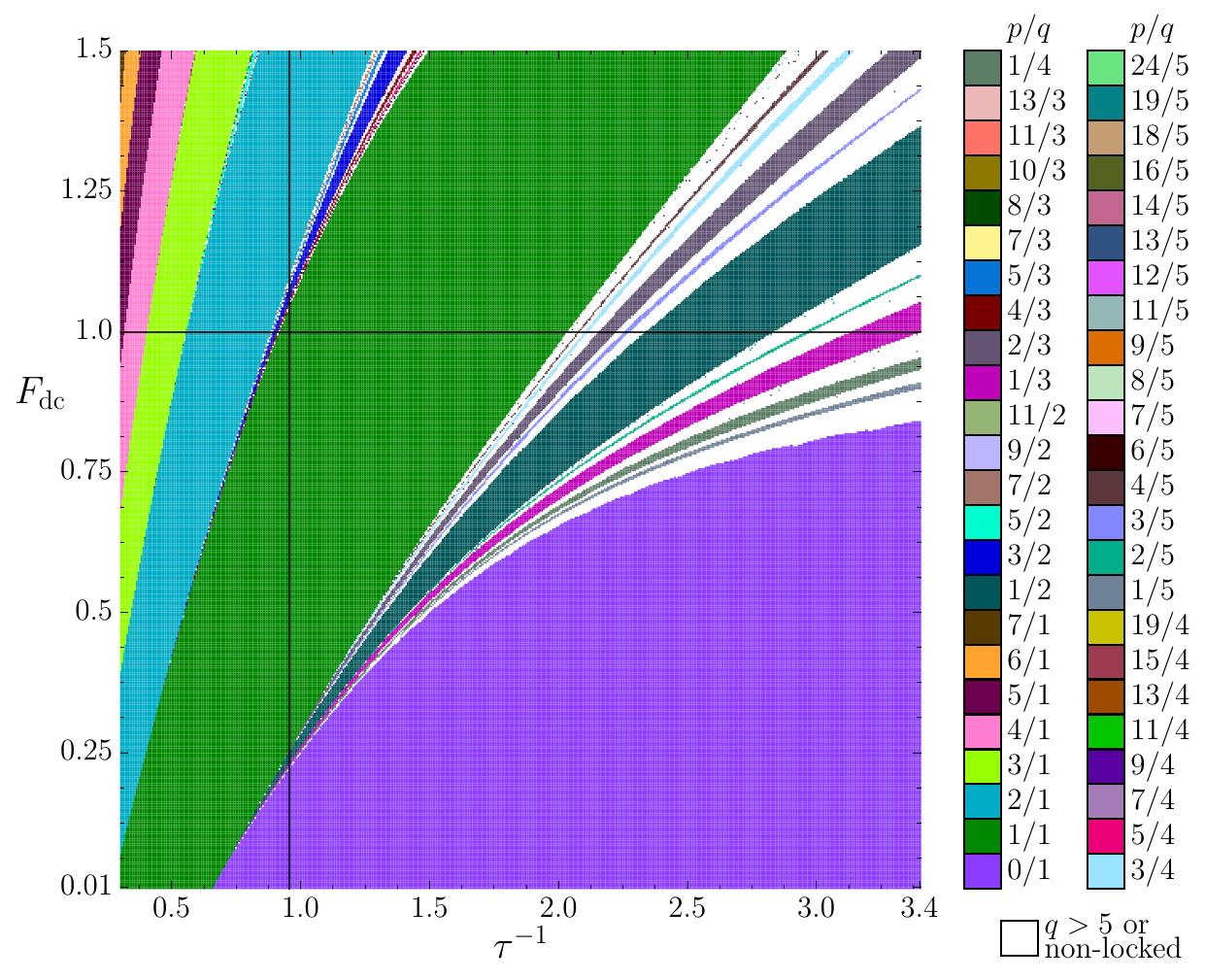}
\caption{Diagram of $p/q$ values specifying phase-locked many-particle
  currents $J_{p,q}$ for hard-sphere diameter $\sigma=0.4\lambda$ in
  dependence of $1/\tau$ and $F_{\rm dc}$ for fixed amplitude $F_{\rm
    ac}=1.5$ of the time periodic driving.  The data were calculated
  from Eqs.~\eqref{eq:q-determ} and \eqref{eq:p-determ} with the
  numerical procedure described after Eq.~\eqref{eq:p-determ}.
  Parameter regions, where modes with $q>5$ can occur, as well as
  aperiodic states not synchronized with the driving, are colored in
  white.  Following the vertical black line in the diagram gives the
  phase-locked modes of currents in the zero-noise limit shown in
  Fig.~2(a) of the main text (variation of $F_{\rm dc}$ at fixed
  $F_{\rm ac}=1.5$ and $1/\tau=3/\pi$). Following the horizontal black
  line in the diagram gives the phase-locked modes of currents in the
  zero-noise limit shown in Fig.~2(b) of the main text (variation of
  $1/\tau$ at fixed $F_{\rm dc}=1$ and $F_{\rm ac}=1.5$).}
\label{fig:mode-diagram}
\end{figure}

Using this method, we can scan the parameter space of the driving and
determine for each $\Gamma$ which mode $p/q$ of phase-locked
many-particle current occurs. This in particular reveals the parameter
region where a certain mode appears.  We found very good agreement
with simulated data.  As a demonstration, we show in
Fig.~\ref{fig:mode-diagram} the diagram of phase-locked current modes
in dependence of $1/\tau$ and $F_{\rm dc}$ for particle diameter
$\sigma=0.4\lambda$ at fixed amplitude $F_{\rm ac}=1.5$ of the
time-periodic driving. Modes are shown up to $q=5$. Areas of $1/\tau$
and $F_{\rm dc}$ values marked in white color contain either
phase-locked modes with $q>5$ or points, where aperiodic
nonsychronized dynamics occurs.

When following the black horizontal line at fixed $F_{\rm dc}=1$ in
Fig.~\ref{fig:mode-diagram}, we can infer the phase-locked current
modes shown in Fig.~2(b) of the main text for $\sigma=0.4\lambda$
(variation of $1/\tau$ at fixed $F_{\rm dc}=1$ and $F_{\rm ac}=1.5$).
For example, the integer Shapiro steps at phase-locked currents
$J_{p,q}=(p/q)\lambda/L\tau$ with $p/q=2/1$ and $p/q=1/1$ appear for
$1/\tau$ in the intervals 0.56-0.87 and 0.92-2.03 in
Fig.~\ref{fig:mode-diagram}, as can be inferred from the intervals
where the horizontal black line runs through the areas marked in
turquoise-blue and green. This is in excellent agreement with the
intervals of $1/\tau$, where the corresponding phase-locked current
modes appear in Fig.~2(b). Likewise, the intervals $1/\tau$, where the
fractional Shapiro steps with $p/q=2/3$, 1/2, 2/5, and 1/3 appear in
Fig.~2(b) are correctly predicted by the mode diagram in
Fig.~\ref{fig:mode-diagram}, see the intervals, where the horizontal
black line runs through respective colored areas.

When following the black vertical line at fixed $1/\tau=3/\pi$ in
Fig.~\ref{fig:mode-diagram}, we can infer the phase-locked current
modes shown in Fig.~2(a) of the main text for $\sigma=0.4\lambda$
(variation of $F_{\rm dc}$ at fixed $F_{\rm ac}=1.5$ and
$1/\tau=3/\pi$).  Comparing the appearance of predicted phase-locked
currents in Fig.~2(a) with the predicted one from
Fig.~\ref{fig:mode-diagram}, we again find excellent agreement.


\begin{thebibliography}{45}%
\makeatletter
\providecommand \@ifxundefined [1]{%
 \@ifx{#1\undefined}
}%
\providecommand \@ifnum [1]{%
 \ifnum #1\expandafter \@firstoftwo
 \else \expandafter \@secondoftwo
 \fi
}%
\providecommand \@ifx [1]{%
 \ifx #1\expandafter \@firstoftwo
 \else \expandafter \@secondoftwo
 \fi
}%
\providecommand \natexlab [1]{#1}%
\providecommand \enquote  [1]{``#1''}%
\providecommand \bibnamefont  [1]{#1}%
\providecommand \bibfnamefont [1]{#1}%
\providecommand \citenamefont [1]{#1}%
\providecommand \href@noop [0]{\@secondoftwo}%
\providecommand \href [0]{\begingroup \@sanitize@url \@href}%
\providecommand \@href[1]{\@@startlink{#1}\@@href}%
\providecommand \@@href[1]{\endgroup#1\@@endlink}%
\providecommand \@sanitize@url [0]{\catcode `\\12\catcode `\$12\catcode
  `\&12\catcode `\#12\catcode `\^12\catcode `\_12\catcode `\%12\relax}%
\providecommand \@@startlink[1]{}%
\providecommand \@@endlink[0]{}%
\providecommand \url  [0]{\begingroup\@sanitize@url \@url }%
\providecommand \@url [1]{\endgroup\@href {#1}{\urlprefix }}%
\providecommand \urlprefix  [0]{URL }%
\providecommand \Eprint [0]{\href }%
\providecommand \doibase [0]{https://doi.org/}%
\providecommand \selectlanguage [0]{\@gobble}%
\providecommand \bibinfo  [0]{\@secondoftwo}%
\providecommand \bibfield  [0]{\@secondoftwo}%
\providecommand \translation [1]{[#1]}%
\providecommand \BibitemOpen [0]{}%
\providecommand \bibitemStop [0]{}%
\providecommand \bibitemNoStop [0]{.\EOS\space}%
\providecommand \EOS [0]{\spacefactor3000\relax}%
\providecommand \BibitemShut  [1]{\csname bibitem#1\endcsname}%
\let\auto@bib@innerbib\@empty
\bibitem [{\citenamefont {Pikovsky}\ \emph {et~al.}(2001)\citenamefont
  {Pikovsky}, \citenamefont {Rosenblum},\ and\ \citenamefont
  {Kurths}}]{Pikovsky/etal:2001}%
  \BibitemOpen
  \bibfield  {author} {\bibinfo {author} {\bibfnamefont {A.}~\bibnamefont
  {Pikovsky}}, \bibinfo {author} {\bibfnamefont {M.}~\bibnamefont
  {Rosenblum}},\ and\ \bibinfo {author} {\bibfnamefont {J.}~\bibnamefont
  {Kurths}},\ }\href {https://doi.org/DOI: 10.1017/CBO9780511755743} {\emph
  {\bibinfo {title} {Synchronization: A Universal Concept in Nonlinear
  Sciences}}},\ Cambridge Nonlinear Science Series\ (\bibinfo  {publisher}
  {Cambridge University Press},\ \bibinfo {address} {Cambridge},\ \bibinfo
  {year} {2001})\BibitemShut {NoStop}%
\bibitem [{\citenamefont {Josephson}(1962)}]{Josephson:1962}%
  \BibitemOpen
  \bibfield  {author} {\bibinfo {author} {\bibfnamefont {B.~D.}\ \bibnamefont
  {Josephson}},\ }\bibfield  {title} {\bibinfo {title} {Possible new effects in
  superconductive tunnelling},\ }\href
  {https://doi.org/https://doi.org/10.1016/0031-9163(62)91369-0} {\bibfield
  {journal} {\bibinfo  {journal} {Phys. Lett.}\ }\textbf {\bibinfo {volume}
  {1}},\ \bibinfo {pages} {251} (\bibinfo {year} {1962})}\BibitemShut {NoStop}%
\bibitem [{\citenamefont {Shapiro}(1963)}]{Shapiro:1963}%
  \BibitemOpen
  \bibfield  {author} {\bibinfo {author} {\bibfnamefont {S.}~\bibnamefont
  {Shapiro}},\ }\bibfield  {title} {\bibinfo {title} {Josephson currents in
  superconducting tunneling: The effect of microwaves and other observations},\
  }\href {https://doi.org/10.1103/PhysRevLett.11.80} {\bibfield  {journal}
  {\bibinfo  {journal} {Phys. Rev. Lett.}\ }\textbf {\bibinfo {volume} {11}},\
  \bibinfo {pages} {80} (\bibinfo {year} {1963})}\BibitemShut {NoStop}%
\bibitem [{\citenamefont {Dinsmore}\ \emph {et~al.}(2008)\citenamefont
  {Dinsmore}, \citenamefont {Bae},\ and\ \citenamefont
  {Bezryadin}}]{Dinsmore/etal:2008}%
  \BibitemOpen
  \bibfield  {author} {\bibinfo {author} {\bibfnamefont {I.}~\bibnamefont
  {Dinsmore}, \bibfnamefont {R.~C.}}, \bibinfo {author} {\bibfnamefont {M.-H.}\
  \bibnamefont {Bae}},\ and\ \bibinfo {author} {\bibfnamefont {A.}~\bibnamefont
  {Bezryadin}},\ }\bibfield  {title} {\bibinfo {title} {Fractional order
  {S}hapiro steps in superconducting nanowires},\ }\href
  {https://doi.org/10.1063/1.3012360} {\bibfield  {journal} {\bibinfo
  {journal} {Appl. Phys. Lett.}\ }\textbf {\bibinfo {volume} {93}},\ \bibinfo
  {pages} {192505} (\bibinfo {year} {2008})}\BibitemShut {NoStop}%
\bibitem [{\citenamefont {Bae}\ \emph {et~al.}(2009)\citenamefont {Bae},
  \citenamefont {Dinsmore}, \citenamefont {Aref}, \citenamefont {Brenner},\
  and\ \citenamefont {Bezryadin}}]{Bae/etal:2009}%
  \BibitemOpen
  \bibfield  {author} {\bibinfo {author} {\bibfnamefont {M.-H.}\ \bibnamefont
  {Bae}}, \bibinfo {author} {\bibfnamefont {R.~C.~I.}\ \bibnamefont
  {Dinsmore}}, \bibinfo {author} {\bibfnamefont {T.}~\bibnamefont {Aref}},
  \bibinfo {author} {\bibfnamefont {M.}~\bibnamefont {Brenner}},\ and\ \bibinfo
  {author} {\bibfnamefont {A.}~\bibnamefont {Bezryadin}},\ }\bibfield  {title}
  {\bibinfo {title} {Current-phase relationship, thermal and quantum phase
  slips in superconducting nanowires made on a scaffold created using adhesive
  tape},\ }\href {https://doi.org/10.1021/nl803894m} {\bibfield  {journal}
  {\bibinfo  {journal} {Nano Lett.}\ }\textbf {\bibinfo {volume} {9}},\
  \bibinfo {pages} {1889} (\bibinfo {year} {2009})}\BibitemShut {NoStop}%
\bibitem [{\citenamefont {Ridderbos}\ \emph {et~al.}(2019)\citenamefont
  {Ridderbos}, \citenamefont {Brauns}, \citenamefont {Li}, \citenamefont
  {Bakkers}, \citenamefont {Brinkman}, \citenamefont {van~der Wiel},\ and\
  \citenamefont {Zwanenburg}}]{Ridderbos/etal:2019}%
  \BibitemOpen
  \bibfield  {author} {\bibinfo {author} {\bibfnamefont {J.}~\bibnamefont
  {Ridderbos}}, \bibinfo {author} {\bibfnamefont {M.}~\bibnamefont {Brauns}},
  \bibinfo {author} {\bibfnamefont {A.}~\bibnamefont {Li}}, \bibinfo {author}
  {\bibfnamefont {E.~P. A.~M.}\ \bibnamefont {Bakkers}}, \bibinfo {author}
  {\bibfnamefont {A.}~\bibnamefont {Brinkman}}, \bibinfo {author}
  {\bibfnamefont {W.~G.}\ \bibnamefont {van~der Wiel}},\ and\ \bibinfo {author}
  {\bibfnamefont {F.~A.}\ \bibnamefont {Zwanenburg}},\ }\bibfield  {title}
  {\bibinfo {title} {Multiple {A}ndreev reflections and {S}hapiro steps in a
  {G}e-{S}i nanowire {J}osephson junction},\ }\href
  {https://doi.org/10.1103/PhysRevMaterials.3.084803} {\bibfield  {journal}
  {\bibinfo  {journal} {Phys. Rev. Mater.}\ }\textbf {\bibinfo {volume} {3}},\
  \bibinfo {pages} {084803} (\bibinfo {year} {2019})}\BibitemShut {NoStop}%
\bibitem [{\citenamefont {Gr{\"u}ner}\ and\ \citenamefont
  {Zettl}(1985)}]{Gruener/Zettl:1985}%
  \BibitemOpen
  \bibfield  {author} {\bibinfo {author} {\bibfnamefont {G.}~\bibnamefont
  {Gr{\"u}ner}}\ and\ \bibinfo {author} {\bibfnamefont {A.}~\bibnamefont
  {Zettl}},\ }\bibfield  {title} {\bibinfo {title} {Charge density wave
  conduction: A novel collective transport phenomenon in solids},\ }\href
  {https://doi.org/https://doi.org/10.1016/0370-1573(85)90073-0} {\bibfield
  {journal} {\bibinfo  {journal} {Phys. Rep.}\ }\textbf {\bibinfo {volume}
  {119}},\ \bibinfo {pages} {117} (\bibinfo {year} {1985})}\BibitemShut
  {NoStop}%
\bibitem [{\citenamefont {Gr\"uner}(1988)}]{Gruener:1988}%
  \BibitemOpen
  \bibfield  {author} {\bibinfo {author} {\bibfnamefont {G.}~\bibnamefont
  {Gr\"uner}},\ }\bibfield  {title} {\bibinfo {title} {The dynamics of
  charge-density waves},\ }\href {https://doi.org/10.1103/RevModPhys.60.1129}
  {\bibfield  {journal} {\bibinfo  {journal} {Rev. Mod. Phys.}\ }\textbf
  {\bibinfo {volume} {60}},\ \bibinfo {pages} {1129} (\bibinfo {year}
  {1988})}\BibitemShut {NoStop}%
\bibitem [{\citenamefont {Thorne}\ \emph {et~al.}(1988)\citenamefont {Thorne},
  \citenamefont {Hubacek}, \citenamefont {Lyons}, \citenamefont {Lyding},\ and\
  \citenamefont {Tucker}}]{Thorne/etal:1988}%
  \BibitemOpen
  \bibfield  {author} {\bibinfo {author} {\bibfnamefont {R.~E.}\ \bibnamefont
  {Thorne}}, \bibinfo {author} {\bibfnamefont {J.~S.}\ \bibnamefont {Hubacek}},
  \bibinfo {author} {\bibfnamefont {W.~G.}\ \bibnamefont {Lyons}}, \bibinfo
  {author} {\bibfnamefont {J.~W.}\ \bibnamefont {Lyding}},\ and\ \bibinfo
  {author} {\bibfnamefont {J.~R.}\ \bibnamefont {Tucker}},\ }\bibfield  {title}
  {\bibinfo {title} {ac-dc interference, complete mode locking, and origin of
  coherent oscillations in sliding charge-density-wave systems},\ }\href
  {https://doi.org/10.1103/PhysRevB.37.10055} {\bibfield  {journal} {\bibinfo
  {journal} {Phys. Rev. B}\ }\textbf {\bibinfo {volume} {37}},\ \bibinfo
  {pages} {10055} (\bibinfo {year} {1988})}\BibitemShut {NoStop}%
\bibitem [{\citenamefont {Hundley}\ and\ \citenamefont
  {Zettl}(1989)}]{Hundley/Zettl:1989}%
  \BibitemOpen
  \bibfield  {author} {\bibinfo {author} {\bibfnamefont {M.~F.}\ \bibnamefont
  {Hundley}}\ and\ \bibinfo {author} {\bibfnamefont {A.}~\bibnamefont
  {Zettl}},\ }\bibfield  {title} {\bibinfo {title} {Noise and ac-dc
  interference phenomena in the charge-density-wave conductor
  $\mathrm{K}_{0.3}\mathrm{MoO}_3$},\ }\href
  {https://doi.org/10.1103/PhysRevB.39.3026} {\bibfield  {journal} {\bibinfo
  {journal} {Phys. Rev. B}\ }\textbf {\bibinfo {volume} {39}},\ \bibinfo
  {pages} {3026} (\bibinfo {year} {1989})}\BibitemShut {NoStop}%
\bibitem [{\citenamefont {Sinchenko}\ and\ \citenamefont
  {Monceau}(2013)}]{Sinchenko/Monceau:2013}%
  \BibitemOpen
  \bibfield  {author} {\bibinfo {author} {\bibfnamefont {A.~A.}\ \bibnamefont
  {Sinchenko}}\ and\ \bibinfo {author} {\bibfnamefont {P.}~\bibnamefont
  {Monceau}},\ }\bibfield  {title} {\bibinfo {title} {Dynamical transport
  properties of $\mathrm{NbSe}_3$ with simultaneous sliding of both
  charge-density waves},\ }\href {https://doi.org/10.1103/PhysRevB.87.045105}
  {\bibfield  {journal} {\bibinfo  {journal} {Phys. Rev. B}\ }\textbf {\bibinfo
  {volume} {87}},\ \bibinfo {pages} {045105} (\bibinfo {year}
  {2013})}\BibitemShut {NoStop}%
\bibitem [{\citenamefont {Reichhardt}\ and\ \citenamefont
  {Reichhardt}(2015)}]{Reichardt/Reichardt:2015}%
  \BibitemOpen
  \bibfield  {author} {\bibinfo {author} {\bibfnamefont {C.}~\bibnamefont
  {Reichhardt}}\ and\ \bibinfo {author} {\bibfnamefont {C.~J.~O.}\ \bibnamefont
  {Reichhardt}},\ }\bibfield  {title} {\bibinfo {title} {Shapiro steps for
  skyrmion motion on a washboard potential with longitudinal and transverse ac
  drives},\ }\href {https://doi.org/10.1103/PhysRevB.92.224432} {\bibfield
  {journal} {\bibinfo  {journal} {Phys. Rev. B}\ }\textbf {\bibinfo {volume}
  {92}},\ \bibinfo {pages} {224432} (\bibinfo {year} {2015})}\BibitemShut
  {NoStop}%
\bibitem [{\citenamefont {Reichhardt}\ and\ \citenamefont
  {Reichhardt}(2017)}]{Reichhardt/Reichhardt:2017a}%
  \BibitemOpen
  \bibfield  {author} {\bibinfo {author} {\bibfnamefont {C.}~\bibnamefont
  {Reichhardt}}\ and\ \bibinfo {author} {\bibfnamefont {C.~J.~O.}\ \bibnamefont
  {Reichhardt}},\ }\bibfield  {title} {\bibinfo {title} {Shapiro spikes and
  negative mobility for skyrmion motion on quasi-one-dimensional periodic
  substrates},\ }\href {https://doi.org/10.1103/PhysRevB.95.014412} {\bibfield
  {journal} {\bibinfo  {journal} {Phys. Rev. B}\ }\textbf {\bibinfo {volume}
  {95}},\ \bibinfo {pages} {014412} (\bibinfo {year} {2017})}\BibitemShut
  {NoStop}%
\bibitem [{\citenamefont {Sato}\ \emph {et~al.}(2020)\citenamefont {Sato},
  \citenamefont {Kikkawa}, \citenamefont {Taguchi}, \citenamefont {Tokura},\
  and\ \citenamefont {Kagawa}}]{Sato/etal:2020}%
  \BibitemOpen
  \bibfield  {author} {\bibinfo {author} {\bibfnamefont {T.}~\bibnamefont
  {Sato}}, \bibinfo {author} {\bibfnamefont {A.}~\bibnamefont {Kikkawa}},
  \bibinfo {author} {\bibfnamefont {Y.}~\bibnamefont {Taguchi}}, \bibinfo
  {author} {\bibfnamefont {Y.}~\bibnamefont {Tokura}},\ and\ \bibinfo {author}
  {\bibfnamefont {F.}~\bibnamefont {Kagawa}},\ }\bibfield  {title} {\bibinfo
  {title} {Mode locking phenomena of the current-induced skyrmion-lattice
  motion in microfabricated {M}n{S}i},\ }\href
  {https://doi.org/10.1103/PhysRevB.102.180411} {\bibfield  {journal} {\bibinfo
   {journal} {Phys. Rev. B}\ }\textbf {\bibinfo {volume} {102}},\ \bibinfo
  {pages} {180411} (\bibinfo {year} {2020})}\BibitemShut {NoStop}%
\bibitem [{\citenamefont {Vizarim}\ \emph {et~al.}(2020)\citenamefont
  {Vizarim}, \citenamefont {Reichhardt}, \citenamefont {Venegas},\ and\
  \citenamefont {Reichhardt}}]{Vizarim/etal:2020}%
  \BibitemOpen
  \bibfield  {author} {\bibinfo {author} {\bibfnamefont {N.~P.}\ \bibnamefont
  {Vizarim}}, \bibinfo {author} {\bibfnamefont {C.~J.~O.}\ \bibnamefont
  {Reichhardt}}, \bibinfo {author} {\bibfnamefont {P.~A.}\ \bibnamefont
  {Venegas}},\ and\ \bibinfo {author} {\bibfnamefont {C.}~\bibnamefont
  {Reichhardt}},\ }\bibfield  {title} {\bibinfo {title} {Skyrmion dynamics and
  transverse mobility: skyrmion hall angle reversal on 2d periodic substrates
  with dc and biharmonic ac drives},\ }\href
  {https://doi.org/10.1140/epjb/e2020-10135-1} {\bibfield  {journal} {\bibinfo
  {journal} {Eur. Phys. J. B}\ }\textbf {\bibinfo {volume} {93}},\ \bibinfo
  {pages} {112} (\bibinfo {year} {2020})}\BibitemShut {NoStop}%
\bibitem [{\citenamefont {Souza}\ \emph {et~al.}(2024)\citenamefont {Souza},
  \citenamefont {Vizarim}, \citenamefont {Reichhardt}, \citenamefont
  {Reichhardt},\ and\ \citenamefont {Venegas}}]{Souza/etal:2024}%
  \BibitemOpen
  \bibfield  {author} {\bibinfo {author} {\bibfnamefont {J.~C.~B.}\
  \bibnamefont {Souza}}, \bibinfo {author} {\bibfnamefont {N.~P.}\ \bibnamefont
  {Vizarim}}, \bibinfo {author} {\bibfnamefont {C.~J.~O.}\ \bibnamefont
  {Reichhardt}}, \bibinfo {author} {\bibfnamefont {C.}~\bibnamefont
  {Reichhardt}},\ and\ \bibinfo {author} {\bibfnamefont {P.~A.}\ \bibnamefont
  {Venegas}},\ }\bibfield  {title} {\bibinfo {title} {Shapiro steps and
  stability of skyrmions interacting with alternating anisotropy under the
  influence of ac and dc drives},\ }\href
  {https://doi.org/10.1103/PhysRevB.110.014406} {\bibfield  {journal} {\bibinfo
   {journal} {Phys. Rev. B}\ }\textbf {\bibinfo {volume} {110}},\ \bibinfo
  {pages} {014406} (\bibinfo {year} {2024})}\BibitemShut {NoStop}%
\bibitem [{\citenamefont {Reichhardt}\ \emph {et~al.}(2000)\citenamefont
  {Reichhardt}, \citenamefont {Scalettar}, \citenamefont {Zim\'anyi},\ and\
  \citenamefont {Gr\o{}nbech-Jensen}}]{Reichardt/etal:1999}%
  \BibitemOpen
  \bibfield  {author} {\bibinfo {author} {\bibfnamefont {C.}~\bibnamefont
  {Reichhardt}}, \bibinfo {author} {\bibfnamefont {R.~T.}\ \bibnamefont
  {Scalettar}}, \bibinfo {author} {\bibfnamefont {G.~T.}\ \bibnamefont
  {Zim\'anyi}},\ and\ \bibinfo {author} {\bibfnamefont {N.}~\bibnamefont
  {Gr\o{}nbech-Jensen}},\ }\bibfield  {title} {\bibinfo {title} {Phase-locking
  of vortex lattices interacting with periodic pinning},\ }\href
  {https://doi.org/10.1103/PhysRevB.61.R11914} {\bibfield  {journal} {\bibinfo
  {journal} {Phys. Rev. B}\ }\textbf {\bibinfo {volume} {61}},\ \bibinfo
  {pages} {R11914} (\bibinfo {year} {2000})}\BibitemShut {NoStop}%
\bibitem [{\citenamefont {Reichhardt}\ and\ \citenamefont
  {Nori}(1999)}]{Reichhardt/Nori:1999}%
  \BibitemOpen
  \bibfield  {author} {\bibinfo {author} {\bibfnamefont {C.}~\bibnamefont
  {Reichhardt}}\ and\ \bibinfo {author} {\bibfnamefont {F.}~\bibnamefont
  {Nori}},\ }\bibfield  {title} {\bibinfo {title} {Phase locking, devil's
  staircases, {F}arey trees, and {A}rnold tongues in driven vortex lattices
  with periodic pinning},\ }\href {https://doi.org/10.1103/PhysRevLett.82.414}
  {\bibfield  {journal} {\bibinfo  {journal} {Phys. Rev. Lett.}\ }\textbf
  {\bibinfo {volume} {82}},\ \bibinfo {pages} {414} (\bibinfo {year}
  {1999})}\BibitemShut {NoStop}%
\bibitem [{\citenamefont {Kolton}\ \emph {et~al.}(2001)\citenamefont {Kolton},
  \citenamefont {Dom\'{\i}nguez},\ and\ \citenamefont
  {Gr\o{}nbech-Jensen}}]{Kolton/etal:2001}%
  \BibitemOpen
  \bibfield  {author} {\bibinfo {author} {\bibfnamefont {A.~B.}\ \bibnamefont
  {Kolton}}, \bibinfo {author} {\bibfnamefont {D.}~\bibnamefont
  {Dom\'{\i}nguez}},\ and\ \bibinfo {author} {\bibfnamefont {N.}~\bibnamefont
  {Gr\o{}nbech-Jensen}},\ }\bibfield  {title} {\bibinfo {title} {Mode locking
  in ac-driven vortex lattices with random pinning},\ }\href
  {https://doi.org/10.1103/PhysRevLett.86.4112} {\bibfield  {journal} {\bibinfo
   {journal} {Phys. Rev. Lett.}\ }\textbf {\bibinfo {volume} {86}},\ \bibinfo
  {pages} {4112} (\bibinfo {year} {2001})}\BibitemShut {NoStop}%
\bibitem [{\citenamefont {Juniper}\ \emph {et~al.}(2015)\citenamefont
  {Juniper}, \citenamefont {Straube}, \citenamefont {Besseling}, \citenamefont
  {Aarts},\ and\ \citenamefont {Dullens}}]{Juniper/etal:2015}%
  \BibitemOpen
  \bibfield  {author} {\bibinfo {author} {\bibfnamefont {M.~P.~N.}\
  \bibnamefont {Juniper}}, \bibinfo {author} {\bibfnamefont {A.~V.}\
  \bibnamefont {Straube}}, \bibinfo {author} {\bibfnamefont {R.}~\bibnamefont
  {Besseling}}, \bibinfo {author} {\bibfnamefont {D.~G. A.~L.}\ \bibnamefont
  {Aarts}},\ and\ \bibinfo {author} {\bibfnamefont {R.~P.~A.}\ \bibnamefont
  {Dullens}},\ }\bibfield  {title} {\bibinfo {title} {Microscopic dynamics of
  synchronization in driven colloids},\ }\href
  {https://doi.org/10.1038/ncomms8187} {\bibfield  {journal} {\bibinfo
  {journal} {Nat. Commun.}\ }\textbf {\bibinfo {volume} {6}},\ \bibinfo {pages}
  {7187} (\bibinfo {year} {2015})}\BibitemShut {NoStop}%
\bibitem [{\citenamefont {Juniper}\ \emph {et~al.}(2017)\citenamefont
  {Juniper}, \citenamefont {Zimmermann}, \citenamefont {Straube}, \citenamefont
  {Besseling}, \citenamefont {Aarts}, \citenamefont {L{\"o}wen},\ and\
  \citenamefont {Dullens}}]{Juniper/etal:2017}%
  \BibitemOpen
  \bibfield  {author} {\bibinfo {author} {\bibfnamefont {M.~P.~N.}\
  \bibnamefont {Juniper}}, \bibinfo {author} {\bibfnamefont {U.}~\bibnamefont
  {Zimmermann}}, \bibinfo {author} {\bibfnamefont {A.~V.}\ \bibnamefont
  {Straube}}, \bibinfo {author} {\bibfnamefont {R.}~\bibnamefont {Besseling}},
  \bibinfo {author} {\bibfnamefont {D.~G. A.~L.}\ \bibnamefont {Aarts}},
  \bibinfo {author} {\bibfnamefont {H.}~\bibnamefont {L{\"o}wen}},\ and\
  \bibinfo {author} {\bibfnamefont {R.~P.~A.}\ \bibnamefont {Dullens}},\
  }\bibfield  {title} {\bibinfo {title} {Dynamic mode locking in a driven
  colloidal system: experiments and theory},\ }\href
  {https://doi.org/10.1088/1367-2630/aa53cd} {\bibfield  {journal} {\bibinfo
  {journal} {New J. Phys.}\ }\textbf {\bibinfo {volume} {19}},\ \bibinfo
  {pages} {013010} (\bibinfo {year} {2017})}\BibitemShut {NoStop}%
\bibitem [{\citenamefont {Tierno}\ and\ \citenamefont
  {Fischer}(2014)}]{Tierno/Fischer:2014}%
  \BibitemOpen
  \bibfield  {author} {\bibinfo {author} {\bibfnamefont {P.}~\bibnamefont
  {Tierno}}\ and\ \bibinfo {author} {\bibfnamefont {T.~M.}\ \bibnamefont
  {Fischer}},\ }\bibfield  {title} {\bibinfo {title} {Excluded volume causes
  integer and fractional plateaus in colloidal ratchet currents},\ }\href
  {https://doi.org/10.1103/PhysRevLett.112.048302} {\bibfield  {journal}
  {\bibinfo  {journal} {Phys. Rev. Lett.}\ }\textbf {\bibinfo {volume} {112}},\
  \bibinfo {pages} {048302} (\bibinfo {year} {2014})}\BibitemShut {NoStop}%
\bibitem [{\citenamefont {Tierno}\ \emph {et~al.}(2014)\citenamefont {Tierno},
  \citenamefont {Johansen},\ and\ \citenamefont {Fischer}}]{Tierno/etal:2014}%
  \BibitemOpen
  \bibfield  {author} {\bibinfo {author} {\bibfnamefont {P.}~\bibnamefont
  {Tierno}}, \bibinfo {author} {\bibfnamefont {T.~H.}\ \bibnamefont
  {Johansen}},\ and\ \bibinfo {author} {\bibfnamefont {T.~M.}\ \bibnamefont
  {Fischer}},\ }\bibfield  {title} {\bibinfo {title} {Fast and rewritable
  colloidal assembly via field synchronized particle swapping},\ }\href
  {https://doi.org/10.1063/1.4874839} {\bibfield  {journal} {\bibinfo
  {journal} {Appl. Phys. Lett.}\ }\textbf {\bibinfo {volume} {104}},\ \bibinfo
  {pages} {174102} (\bibinfo {year} {2014})}\BibitemShut {NoStop}%
\bibitem [{\citenamefont {Bohlein}\ \emph {et~al.}(2012)\citenamefont
  {Bohlein}, \citenamefont {Mikhael},\ and\ \citenamefont
  {Bechinger}}]{Bohlein/etal:2012}%
  \BibitemOpen
  \bibfield  {author} {\bibinfo {author} {\bibfnamefont {T.}~\bibnamefont
  {Bohlein}}, \bibinfo {author} {\bibfnamefont {J.}~\bibnamefont {Mikhael}},\
  and\ \bibinfo {author} {\bibfnamefont {C.}~\bibnamefont {Bechinger}},\
  }\bibfield  {title} {\bibinfo {title} {Observation of kinks and antikinks in
  colloidal monolayers driven across ordered surfaces},\ }\href
  {https://doi.org/10.1038/nmat3204} {\bibfield  {journal} {\bibinfo  {journal}
  {Nat. Mater.}\ }\textbf {\bibinfo {volume} {11}},\ \bibinfo {pages} {126}
  (\bibinfo {year} {2012})}\BibitemShut {NoStop}%
\bibitem [{\citenamefont {Bohlein}\ and\ \citenamefont
  {Bechinger}(2012)}]{Bohlein/Bechinger:2012}%
  \BibitemOpen
  \bibfield  {author} {\bibinfo {author} {\bibfnamefont {T.}~\bibnamefont
  {Bohlein}}\ and\ \bibinfo {author} {\bibfnamefont {C.}~\bibnamefont
  {Bechinger}},\ }\bibfield  {title} {\bibinfo {title} {Experimental
  observation of directional locking and dynamical ordering of colloidal
  monolayers driven across quasiperiodic substrates},\ }\href
  {https://doi.org/10.1103/PhysRevLett.109.058301} {\bibfield  {journal}
  {\bibinfo  {journal} {Phys. Rev. Lett.}\ }\textbf {\bibinfo {volume} {109}},\
  \bibinfo {pages} {058301} (\bibinfo {year} {2012})}\BibitemShut {NoStop}%
\bibitem [{\citenamefont {Brazda}\ \emph {et~al.}(2017)\citenamefont {Brazda},
  \citenamefont {July},\ and\ \citenamefont {Bechinger}}]{Brazda/etal:2017}%
  \BibitemOpen
  \bibfield  {author} {\bibinfo {author} {\bibfnamefont {T.}~\bibnamefont
  {Brazda}}, \bibinfo {author} {\bibfnamefont {C.}~\bibnamefont {July}},\ and\
  \bibinfo {author} {\bibfnamefont {C.}~\bibnamefont {Bechinger}},\ }\bibfield
  {title} {\bibinfo {title} {Experimental observation of {S}hapiro-steps in
  colloidal monolayers driven across time-dependent substrate potentials},\
  }\href {https://doi.org/10.1039/C7SM00393E} {\bibfield  {journal} {\bibinfo
  {journal} {Soft Matter}\ }\textbf {\bibinfo {volume} {13}},\ \bibinfo {pages}
  {4024} (\bibinfo {year} {2017})}\BibitemShut {NoStop}%
\bibitem [{\citenamefont {Brazda}\ \emph {et~al.}(2018)\citenamefont {Brazda},
  \citenamefont {Silva}, \citenamefont {Manini}, \citenamefont {Vanossi},
  \citenamefont {Guerra}, \citenamefont {Tosatti},\ and\ \citenamefont
  {Bechinger}}]{Brazda/etal:2018}%
  \BibitemOpen
  \bibfield  {author} {\bibinfo {author} {\bibfnamefont {T.}~\bibnamefont
  {Brazda}}, \bibinfo {author} {\bibfnamefont {A.}~\bibnamefont {Silva}},
  \bibinfo {author} {\bibfnamefont {N.}~\bibnamefont {Manini}}, \bibinfo
  {author} {\bibfnamefont {A.}~\bibnamefont {Vanossi}}, \bibinfo {author}
  {\bibfnamefont {R.}~\bibnamefont {Guerra}}, \bibinfo {author} {\bibfnamefont
  {E.}~\bibnamefont {Tosatti}},\ and\ \bibinfo {author} {\bibfnamefont
  {C.}~\bibnamefont {Bechinger}},\ }\bibfield  {title} {\bibinfo {title}
  {Experimental observation of the {Aubry} transition in two-dimensional
  colloidal monolayers},\ }\href {https://doi.org/10.1103/PhysRevX.8.011050}
  {\bibfield  {journal} {\bibinfo  {journal} {Phys. Rev. X}\ }\textbf {\bibinfo
  {volume} {8}},\ \bibinfo {pages} {011050} (\bibinfo {year}
  {2018})}\BibitemShut {NoStop}%
\bibitem [{\citenamefont {Vanossi}\ \emph {et~al.}(2020)\citenamefont
  {Vanossi}, \citenamefont {Bechinger},\ and\ \citenamefont
  {Urbakh}}]{Vanossi/etal:2020}%
  \BibitemOpen
  \bibfield  {author} {\bibinfo {author} {\bibfnamefont {A.}~\bibnamefont
  {Vanossi}}, \bibinfo {author} {\bibfnamefont {C.}~\bibnamefont {Bechinger}},\
  and\ \bibinfo {author} {\bibfnamefont {M.}~\bibnamefont {Urbakh}},\
  }\bibfield  {title} {\bibinfo {title} {Structural lubricity in soft and hard
  matter systems},\ }\href {https://doi.org/10.1038/s41467-020-18429-1}
  {\bibfield  {journal} {\bibinfo  {journal} {Nat. Commun.}\ }\textbf {\bibinfo
  {volume} {11}},\ \bibinfo {pages} {4657} (\bibinfo {year}
  {2020})}\BibitemShut {NoStop}%
\bibitem [{\citenamefont {Waldram}\ and\ \citenamefont
  {Wu}(1982)}]{Waldram/Wu:1982}%
  \BibitemOpen
  \bibfield  {author} {\bibinfo {author} {\bibfnamefont {J.~R.}\ \bibnamefont
  {Waldram}}\ and\ \bibinfo {author} {\bibfnamefont {P.~H.}\ \bibnamefont
  {Wu}},\ }\bibfield  {title} {\bibinfo {title} {An alternative analysis of the
  nonlinear equations of the current-driven {J}osephson junction},\ }\href
  {https://doi.org/10.1007/BF00683738} {\bibfield  {journal} {\bibinfo
  {journal} {J. Low Temp. Phys.}\ }\textbf {\bibinfo {volume} {47}},\ \bibinfo
  {pages} {363} (\bibinfo {year} {1982})}\BibitemShut {NoStop}%
\bibitem [{\citenamefont {Antonov}\ \emph
  {et~al.}(2022{\natexlab{a}})\citenamefont {Antonov}, \citenamefont {Ryabov},\
  and\ \citenamefont {Maass}}]{Antonov/etal:2022a}%
  \BibitemOpen
  \bibfield  {author} {\bibinfo {author} {\bibfnamefont {A.~P.}\ \bibnamefont
  {Antonov}}, \bibinfo {author} {\bibfnamefont {A.}~\bibnamefont {Ryabov}},\
  and\ \bibinfo {author} {\bibfnamefont {P.}~\bibnamefont {Maass}},\ }\bibfield
   {title} {\bibinfo {title} {Solitons in overdamped {B}rownian dynamics},\
  }\href {https://doi.org/10.1103/PhysRevLett.129.080601} {\bibfield  {journal}
  {\bibinfo  {journal} {Phys. Rev. Lett.}\ }\textbf {\bibinfo {volume} {129}},\
  \bibinfo {pages} {080601} (\bibinfo {year} {2022}{\natexlab{a}})}\BibitemShut
  {NoStop}%
\bibitem [{\citenamefont {Cereceda-L\'opez}\ \emph {et~al.}(2023)\citenamefont
  {Cereceda-L\'opez}, \citenamefont {Antonov}, \citenamefont {Ryabov},
  \citenamefont {Maass},\ and\ \citenamefont
  {Tierno}}]{Cereceda-Lopez/etal:2023}%
  \BibitemOpen
  \bibfield  {author} {\bibinfo {author} {\bibfnamefont {E.}~\bibnamefont
  {Cereceda-L\'opez}}, \bibinfo {author} {\bibfnamefont {A.~P.}\ \bibnamefont
  {Antonov}}, \bibinfo {author} {\bibfnamefont {A.}~\bibnamefont {Ryabov}},
  \bibinfo {author} {\bibfnamefont {P.}~\bibnamefont {Maass}},\ and\ \bibinfo
  {author} {\bibfnamefont {P.}~\bibnamefont {Tierno}},\ }\bibfield  {title}
  {\bibinfo {title} {Overcrowding induces fast colloidal solitons in a slowly
  rotating potential landscape},\ }\href
  {https://doi.org/10.1038/s41467-023-41989-x} {\bibfield  {journal} {\bibinfo
  {journal} {Nat. Commun.}\ }\textbf {\bibinfo {volume} {14}},\ \bibinfo
  {pages} {6448} (\bibinfo {year} {2023})}\BibitemShut {NoStop}%
\bibitem [{\citenamefont {Cereceda-L\'opez}\ \emph {et~al.}(2021)\citenamefont
  {Cereceda-L\'opez}, \citenamefont {Lips}, \citenamefont {Ortiz-Ambriz},
  \citenamefont {Ryabov}, \citenamefont {Maass},\ and\ \citenamefont
  {Tierno}}]{Cereceda-Lopez/etal:2021}%
  \BibitemOpen
  \bibfield  {author} {\bibinfo {author} {\bibfnamefont {E.}~\bibnamefont
  {Cereceda-L\'opez}}, \bibinfo {author} {\bibfnamefont {D.}~\bibnamefont
  {Lips}}, \bibinfo {author} {\bibfnamefont {A.}~\bibnamefont {Ortiz-Ambriz}},
  \bibinfo {author} {\bibfnamefont {A.}~\bibnamefont {Ryabov}}, \bibinfo
  {author} {\bibfnamefont {P.}~\bibnamefont {Maass}},\ and\ \bibinfo {author}
  {\bibfnamefont {P.}~\bibnamefont {Tierno}},\ }\bibfield  {title} {\bibinfo
  {title} {Hydrodynamic interactions can induce jamming in flow-driven
  systems},\ }\href {https://doi.org/10.1103/PhysRevLett.127.214501} {\bibfield
   {journal} {\bibinfo  {journal} {Phys. Rev. Lett.}\ }\textbf {\bibinfo
  {volume} {127}},\ \bibinfo {pages} {214501} (\bibinfo {year}
  {2021})}\BibitemShut {NoStop}%
\bibitem [{\citenamefont {Antonov}\ \emph {et~al.}(2024)\citenamefont
  {Antonov}, \citenamefont {Ryabov},\ and\ \citenamefont
  {Maass}}]{Antonov/etal:2024}%
  \BibitemOpen
  \bibfield  {author} {\bibinfo {author} {\bibfnamefont {A.~P.}\ \bibnamefont
  {Antonov}}, \bibinfo {author} {\bibfnamefont {A.}~\bibnamefont {Ryabov}},\
  and\ \bibinfo {author} {\bibfnamefont {P.}~\bibnamefont {Maass}},\ }\bibfield
   {title} {\bibinfo {title} {Solitary cluster waves in periodic potentials:
  Formation, propagation, and soliton-mediated particle transport},\ }\href
  {https://doi.org/https://doi.org/10.1016/j.chaos.2024.115079} {\bibfield
  {journal} {\bibinfo  {journal} {Chaos, Solitons \& Fractals}\ }\textbf
  {\bibinfo {volume} {185}},\ \bibinfo {pages} {115079} (\bibinfo {year}
  {2024})}\BibitemShut {NoStop}%
\bibitem [{\citenamefont {Kautz}(1996)}]{Kautz:1996}%
  \BibitemOpen
  \bibfield  {author} {\bibinfo {author} {\bibfnamefont {R.~L.}\ \bibnamefont
  {Kautz}},\ }\bibfield  {title} {\bibinfo {title} {Noise, chaos, and the
  {J}osephson voltage standard},\ }\href
  {https://doi.org/10.1088/0034-4885/59/8/001} {\bibfield  {journal} {\bibinfo
  {journal} {Rep. Prog. Phys.}\ }\textbf {\bibinfo {volume} {59}},\ \bibinfo
  {pages} {935} (\bibinfo {year} {1996})}\BibitemShut {NoStop}%
\bibitem [{\citenamefont {Teki\ifmmode~\acute{c}\else \'{c}\fi{}}\ and\
  \citenamefont {Hu}(2008)}]{Tekic/Hu:2008}%
  \BibitemOpen
  \bibfield  {author} {\bibinfo {author} {\bibfnamefont {J.}~\bibnamefont
  {Teki\ifmmode~\acute{c}\else \'{c}\fi{}}}\ and\ \bibinfo {author}
  {\bibfnamefont {B.}~\bibnamefont {Hu}},\ }\bibfield  {title} {\bibinfo
  {title} {Noise-induced {B}essel-like oscillations of {S}hapiro steps with the
  period of the ac force},\ }\href {https://doi.org/10.1103/PhysRevB.78.104305}
  {\bibfield  {journal} {\bibinfo  {journal} {Phys. Rev. B}\ }\textbf {\bibinfo
  {volume} {78}},\ \bibinfo {pages} {104305} (\bibinfo {year}
  {2008})}\BibitemShut {NoStop}%
\bibitem [{\citenamefont {Mali}\ \emph {et~al.}(2012)\citenamefont {Mali},
  \citenamefont {Teki\'c}, \citenamefont {Ivi\'c},\ and\ \citenamefont
  {Panti\'c}}]{Mali/etal:2012}%
  \BibitemOpen
  \bibfield  {author} {\bibinfo {author} {\bibfnamefont {P.}~\bibnamefont
  {Mali}}, \bibinfo {author} {\bibfnamefont {J.}~\bibnamefont {Teki\'c}},
  \bibinfo {author} {\bibfnamefont {Z.}~\bibnamefont {Ivi\'c}},\ and\ \bibinfo
  {author} {\bibfnamefont {M.}~\bibnamefont {Panti\'c}},\ }\bibfield  {title}
  {\bibinfo {title} {Effects of noise on interference phenomena in the presence
  of subharmonic {S}hapiro steps},\ }\href
  {https://doi.org/10.1103/PhysRevE.86.046209} {\bibfield  {journal} {\bibinfo
  {journal} {Phys. Rev. E}\ }\textbf {\bibinfo {volume} {86}},\ \bibinfo
  {pages} {046209} (\bibinfo {year} {2012})}\BibitemShut {NoStop}%
\bibitem [{\citenamefont {Mali}\ \emph {et~al.}(2020)\citenamefont {Mali},
  \citenamefont {\v{S}akota}, \citenamefont {Teki\'c}, \citenamefont
  {Rado\v{s}evi\'c}, \citenamefont {Panti\'c},\ and\ \citenamefont
  {Pavkov-Hrvojevi\'c}}]{Mali/etal:2020}%
  \BibitemOpen
  \bibfield  {author} {\bibinfo {author} {\bibfnamefont {P.}~\bibnamefont
  {Mali}}, \bibinfo {author} {\bibfnamefont {A.}~\bibnamefont {\v{S}akota}},
  \bibinfo {author} {\bibfnamefont {J.}~\bibnamefont {Teki\'c}}, \bibinfo
  {author} {\bibfnamefont {S.}~\bibnamefont {Rado\v{s}evi\'c}}, \bibinfo
  {author} {\bibfnamefont {M.}~\bibnamefont {Panti\'c}},\ and\ \bibinfo
  {author} {\bibfnamefont {M.}~\bibnamefont {Pavkov-Hrvojevi\'c}},\ }\bibfield
  {title} {\bibinfo {title} {Complexity of {S}hapiro steps},\ }\href
  {https://doi.org/10.1103/PhysRevE.101.032203} {\bibfield  {journal} {\bibinfo
   {journal} {Phys. Rev. E}\ }\textbf {\bibinfo {volume} {101}},\ \bibinfo
  {pages} {032203} (\bibinfo {year} {2020})}\BibitemShut {NoStop}%
\bibitem [{\citenamefont {Juniper}\ \emph {et~al.}(2016)\citenamefont
  {Juniper}, \citenamefont {Straube}, \citenamefont {Aarts},\ and\
  \citenamefont {Dullens}}]{Juniper/etal:2016}%
  \BibitemOpen
  \bibfield  {author} {\bibinfo {author} {\bibfnamefont {M.~P.~N.}\
  \bibnamefont {Juniper}}, \bibinfo {author} {\bibfnamefont {A.~V.}\
  \bibnamefont {Straube}}, \bibinfo {author} {\bibfnamefont {D.~G. A.~L.}\
  \bibnamefont {Aarts}},\ and\ \bibinfo {author} {\bibfnamefont {R.~P.~A.}\
  \bibnamefont {Dullens}},\ }\bibfield  {title} {\bibinfo {title} {Colloidal
  particles driven across periodic optical-potential-energy landscapes},\
  }\href {https://doi.org/10.1103/PhysRevE.93.012608} {\bibfield  {journal}
  {\bibinfo  {journal} {Phys. Rev. E}\ }\textbf {\bibinfo {volume} {93}},\
  \bibinfo {pages} {012608} (\bibinfo {year} {2016})}\BibitemShut {NoStop}%
\bibitem [{\citenamefont {Paronuzzi~Ticco}\ \emph {et~al.}(2016)\citenamefont
  {Paronuzzi~Ticco}, \citenamefont {Fornasier}, \citenamefont {Manini},
  \citenamefont {Santoro}, \citenamefont {Tosatti},\ and\ \citenamefont
  {Vanossi}}]{Paronozzi-Ticco/etal:2016}%
  \BibitemOpen
  \bibfield  {author} {\bibinfo {author} {\bibfnamefont {S.~V.}\ \bibnamefont
  {Paronuzzi~Ticco}}, \bibinfo {author} {\bibfnamefont {G.}~\bibnamefont
  {Fornasier}}, \bibinfo {author} {\bibfnamefont {N.}~\bibnamefont {Manini}},
  \bibinfo {author} {\bibfnamefont {G.~E.}\ \bibnamefont {Santoro}}, \bibinfo
  {author} {\bibfnamefont {E.}~\bibnamefont {Tosatti}},\ and\ \bibinfo {author}
  {\bibfnamefont {A.}~\bibnamefont {Vanossi}},\ }\bibfield  {title} {\bibinfo
  {title} {Subharmonic {S}hapiro steps of sliding colloidal monolayers in
  optical lattices},\ }\href {https://doi.org/10.1088/0953-8984/28/13/134006}
  {\bibfield  {journal} {\bibinfo  {journal} {J. Phys. Condens. Matter}\
  }\textbf {\bibinfo {volume} {28}},\ \bibinfo {pages} {134006} (\bibinfo
  {year} {2016})}\BibitemShut {NoStop}%
\bibitem [{\citenamefont {Antonov}\ \emph
  {et~al.}(2022{\natexlab{b}})\citenamefont {Antonov}, \citenamefont
  {Schweers}, \citenamefont {Ryabov},\ and\ \citenamefont
  {Maass}}]{Antonov/etal:2022c}%
  \BibitemOpen
  \bibfield  {author} {\bibinfo {author} {\bibfnamefont {A.~P.}\ \bibnamefont
  {Antonov}}, \bibinfo {author} {\bibfnamefont {S.}~\bibnamefont {Schweers}},
  \bibinfo {author} {\bibfnamefont {A.}~\bibnamefont {Ryabov}},\ and\ \bibinfo
  {author} {\bibfnamefont {P.}~\bibnamefont {Maass}},\ }\bibfield  {title}
  {\bibinfo {title} {Brownian dynamics simulations of hard rods in external
  fields and with contact interactions},\ }\href
  {https://doi.org/10.1103/PhysRevE.106.054606} {\bibfield  {journal} {\bibinfo
   {journal} {Phys. Rev. E}\ }\textbf {\bibinfo {volume} {106}},\ \bibinfo
  {pages} {054606} (\bibinfo {year} {2022}{\natexlab{b}})}\BibitemShut
  {NoStop}%
\bibitem [{\citenamefont {Antonov}\ \emph {et~al.}(2025)\citenamefont
  {Antonov}, \citenamefont {Schweers}, \citenamefont {Ryabov},\ and\
  \citenamefont {Maass}}]{Antonov/etal:2025}%
  \BibitemOpen
  \bibfield  {author} {\bibinfo {author} {\bibfnamefont {A.~P.}\ \bibnamefont
  {Antonov}}, \bibinfo {author} {\bibfnamefont {S.}~\bibnamefont {Schweers}},
  \bibinfo {author} {\bibfnamefont {A.}~\bibnamefont {Ryabov}},\ and\ \bibinfo
  {author} {\bibfnamefont {P.}~\bibnamefont {Maass}},\ }\bibfield  {title}
  {\bibinfo {title} {Fast brownian cluster dynamics},\ }\href
  {https://doi.org/https://doi.org/10.1016/j.cpc.2024.109474} {\bibfield
  {journal} {\bibinfo  {journal} {Comput. Phys. Commun.}\ }\textbf {\bibinfo
  {volume} {309}},\ \bibinfo {pages} {109474} (\bibinfo {year}
  {2025})}\BibitemShut {NoStop}%
\bibitem [{\citenamefont {Strating}(1999)}]{Strating:1999}%
  \BibitemOpen
  \bibfield  {author} {\bibinfo {author} {\bibfnamefont {P.}~\bibnamefont
  {Strating}},\ }\bibfield  {title} {\bibinfo {title} {Brownian dynamics
  simulation of a hard-sphere suspension},\ }\href
  {https://doi.org/10.1103/PhysRevE.59.2175} {\bibfield  {journal} {\bibinfo
  {journal} {Phys. Rev. E}\ }\textbf {\bibinfo {volume} {59}},\ \bibinfo
  {pages} {2175} (\bibinfo {year} {1999})}\BibitemShut {NoStop}%
\bibitem [{\citenamefont {Scala}(2012)}]{Scala:2012}%
  \BibitemOpen
  \bibfield  {author} {\bibinfo {author} {\bibfnamefont {A.}~\bibnamefont
  {Scala}},\ }\bibfield  {title} {\bibinfo {title} {Event-driven {L}angevin
  simulations of hard spheres},\ }\href
  {https://doi.org/10.1103/PhysRevE.86.026709} {\bibfield  {journal} {\bibinfo
  {journal} {Phys. Rev. E}\ }\textbf {\bibinfo {volume} {86}},\ \bibinfo
  {pages} {026709} (\bibinfo {year} {2012})}\BibitemShut {NoStop}%
\bibitem [{\citenamefont {Strogatz}(2015)}]{Strogatz:2015}%
  \BibitemOpen
  \bibfield  {author} {\bibinfo {author} {\bibfnamefont {S.~H.}\ \bibnamefont
  {Strogatz}},\ }\href {https://doi.org/https://doi.org/10.1201/9780429492563}
  {\emph {\bibinfo {title} {Nonlinear Dynamics and Chaos: With Applications to
  Physics, Biology, Chemistry, and Engineering (2nd ed.)}}}\ (\bibinfo
  {publisher} {CRC Press, Boca Raton},\ \bibinfo {year} {2015})\BibitemShut
  {NoStop}%
\bibitem [{\citenamefont {Press}\ \emph {et~al.}(2007)\citenamefont {Press},
  \citenamefont {Teukolsky}, \citenamefont {Vetterling},\ and\ \citenamefont
  {Flannery}}]{Press/etal:2007}%
  \BibitemOpen
  \bibfield  {author} {\bibinfo {author} {\bibfnamefont {W.~H.}\ \bibnamefont
  {Press}}, \bibinfo {author} {\bibfnamefont {S.~A.}\ \bibnamefont
  {Teukolsky}}, \bibinfo {author} {\bibfnamefont {W.~T.}\ \bibnamefont
  {Vetterling}},\ and\ \bibinfo {author} {\bibfnamefont {B.~P.}\ \bibnamefont
  {Flannery}},\ }\href@noop {} {\emph {\bibinfo {title} {Numerical recipes in
  C: the art of scientific computing}}},\ \bibinfo {edition} {3rd}\ ed.\
  (\bibinfo  {publisher} {Cambridge University Press},\ \bibinfo {address}
  {Cambridge, UK},\ \bibinfo {year} {2007})\BibitemShut {NoStop}%
\end{thebibliography}

\end{document}